\documentclass[a4paper,twocolumn,11pt,accepted=2024-11-21]{quantumarticle}
\pdfoutput=1
\usepackage{float}
\usepackage{amsmath}
\usepackage{amssymb}
\usepackage{mathtools}
\usepackage{amsfonts}
\usepackage{hyperref}
\usepackage{amsthm}
\usepackage{bm}
\usepackage{braket}
\usepackage{cases}
\usepackage{comment}
\usepackage{graphicx}
\usepackage[numbers]{natbib}
\usepackage{color}

\makeatletter
  \renewcommand\@upn{\itshape}
\makeatother
\newtheorem{theorem}{Theorem}
\newtheorem{lemma}{Lemma}

\newtheorem{definition}{Definition}
\newtheorem{remark}{Remark}

\newcommand{\diag}{\mathrm{diag}}

\def\({\left(}

\def\){\right)}

\begin{document}

\title{Tight concentration inequalities for quantum adversarial setups exploiting permutation symmetry}
\author{Takaya Matsuura}
 \email{takaya.matsuura@riken.jp}
 \orcid{0000-0003-4164-4307}
 \affiliation{Centre for Quantum Computation \& Communication Technology,
 School of Science, RMIT University, Melbourne VIC 3000, Australia} 
 \affiliation{RIKEN Center for Quantum Computing (RQC), Hirosawa 2-1, Wako, Saitama 351-0198, Japan}
 \author{Shinichiro Yamano}
 \orcid{0000-0003-3245-5344}
 \affiliation{Department of Applied Physics, Graduate School of Engineering, The University of Tokyo, 7-3-1 Hongo, Bunkyo-ku, Tokyo 113-8656, Japan}
 \author{Yui Kuramochi}
 \orcid{0000-0003-0512-5446}
 \affiliation{Department of Physics, Faculty of Science, Kyushu University, 744 Motooka, Nishi-ku, Fukuoka, Japan}
\author{Toshihiko Sasaki}
\orcid{0000-0003-0745-6791}
 \affiliation{Department of Applied Physics, Graduate School of Engineering, The University of Tokyo, 7-3-1 Hongo, Bunkyo-ku, Tokyo 113-8656, Japan} 
 \affiliation{Photon Science Center, Graduate School of Engineering, The University of Tokyo, 7-3-1 Hongo, Bunkyo-ku, Tokyo 113-8656, Japan} 
\author{Masato Koashi}
\orcid{0000-0002-4518-1461}
 \affiliation{Department of Applied Physics, Graduate School of Engineering, The University of Tokyo, 7-3-1 Hongo, Bunkyo-ku, Tokyo 113-8656, Japan} 
 \affiliation{Photon Science Center, Graduate School of Engineering, The University of Tokyo, 7-3-1 Hongo, Bunkyo-ku, Tokyo 113-8656, Japan}

 \begin{abstract}
    We developed new concentration inequalities for a quantum state on an $N$-qudit system or measurement outcomes on it that apply to an adversarial setup, where an adversary prepares the quantum state.  Our one-sided concentration inequalities for a quantum state require the $N$-qudit system to be permutation invariant and are thus de-Finetti type, but they are tighter than the one previously obtained.  We show that the bound can further be tightened if each qudit system has an additional symmetry.  Furthermore, our concentration inequality for the outcomes of independent and identical measurements on an $N$-qudit quantum system has no assumption on the adversarial quantum state and is much tighter than the conventional one obtained through Azuma's inequality.  We numerically demonstrate the tightness of our bounds in simple quantum information processing tasks.
 \end{abstract}

\maketitle

\section{Introduction and the main results}
\subsection{General introduction}
A concentration inequality is a bound on the probability of obtaining atypical events, i.e., events that are far from the expectation.  It is essential in lots of information-theoretic primitives such as cryptography and error correction since it gives a failure probability in anticipating that the observed values are close to the expectation values. 
The same applies to quantum information theory; we need concentration inequalities to obtain an upper bound on the probability of atypical events in quantum information protocols.
Evaluation of the probability of atypical events may be easy for independent and identically distributed (i.i.d.)~quantum states.  There, one can apply Chernoff-Hoeffding theorem \cite{Hoeffding1963} or Sanov's theorem \cite{Sanov1957, Csiszar1984} to obtain an upper bound on the deviation from the expectation value for an additive observable allowing a small failure probability.
For certain applications, however, one even needs a concentration inequality that applies to an adversarial setup.  This includes quantum key distribution (QKD), verifiable blind quantum computation~\cite{Takeuchi2018}, and verification and learning of a quantum state prepared by an adversary~\cite{Fawzi2024}, where the underlying statistics may be governed by an adversary.

In the field of QKD, for example, proving the security against collective attacks in a finite number of communication rounds is relatively easy, and many QKD protocols have been proved secure at least against collective attacks even in the finite-size regime.  The situation gets involved when considering the security against general attacks.  
There are two mainstreams in the way of proving the security against general attacks in finite-size regimes.  One of them is called the phase-error correction method \cite{Shor2000, Lo1999, Koashi2009, matsuura2023digital}, which is the main focus of this paper, and evaluates the number of phase error patterns or a phase error rate while allowing an exponentially small failure probability.  Correcting the phase error with a failure probability $\epsilon$ ensures $\sim \epsilon^{1/2}$-secrecy of the protocol against the adversary.
In Ref.~\cite{Tamaki2003}, it was shown that in the phase-error correction method, imposing a permutation invariance to the protocol either actively or passively reduces the problem of general attacks to the collective attacks at the cost of increasing the failure probability $\epsilon$ by a polynomial factor.  This implies that the error exponent of the failure probability of the phase error correction is the same against collective attacks and against general attacks. 
After that, a distinct approach using Azuma's inequality, which is based on sequentiality of measurements instead of permutation invariance, was proposed \cite{Boileau2005}.  Since Azuma's inequality is easier to evaluate, it prevails in the QKD community.
At the cost of its simplicity, the error exponent of the failure probability when using Azuma's inequality is always less than that of the collective attack case.  
Intensive research has been made on tightening Azuma's inequality with additional information~\cite{refined_Azuma_ineq1, refined_Azuma_ineq2, Kato2020} including the recently found one with unconfirmed knowledge~\cite{Kato2020}, but it is still looser than the Sanov-type inequalities~\cite{Sanov1957, Csiszar1984} that apply to the i.i.d.~random variables.
Note that in another approach to the security proofs based on the leftover hash lemma against a quantum adversary \cite{Renner2008}, a reduction of the security against general attacks to that against collective attacks by exploiting the permutation symmetry has also been studied under the name of de Finetti reduction and the post-selection technique \cite{Caves2002, Koenig2005, Christandl2007, Renner2007, Renner2008, Christandl2009, Fawzi2015}.  The problems with these approaches were that the reduction was not very straightforward or that they led to worse performance.  Then, the community gradually shifted to an alternative method that uses more recently developed entropy accumulation theorem~\cite{Dupuis2020, Metger2022, Metger2023}.  Interestingly, both Azuma's inequality and the entropy accumulation utilize sequential conditionings by previous measurement outcomes, yet they may be looser than the bound that applies to the i.i.d.~case.

In this paper, we revisit the idea based on permutation invariance and derive concentration inequalities through the reduction to the i.i.d.~scenario. 
The first case we consider is when a quantum state on a multi-qudit system is invariant under the permutation of the subsystems possibly with additional symmetry on each subsystem.  
We show that whenever an outcome of the quantum measurement has a small appearance probability for any i.i.d.~quantum state, it also has a small appearance probability for these permutation-invariant states up to an explicit factor polynomial in the number of subsystems.  This can be proved by observing that any permutation-invariant quantum state is bounded from above by the mixture of i.i.d.~quantum states up to a polynomial factor, i.e., we have the following.
\begin{theorem}[Informal]
    Let $\rho_{A^n}$ be a permutation-invariant state on ${\cal H}_{A}^{\otimes n}$ with $\dim{\cal H}_A =d$.  Then, we have 
    \begin{equation}
        \rho_{A^n} \leq f_q(n,d)\; [\text{Mixture of i.i.d.~states}],
    \end{equation}
    where $f_q(n,d)= {\cal O}(n^{(d^2-1)/2})$ for a large $n$.
\end{theorem}
Compared to the known results \cite{Renner2008, Christandl2009}, we tightly evaluate the polynomial prefactor and improve the bound for a general permutation-invariant state.  We also obtain a bound that explicitly depends on the irreducible representation space of the permutation group.  In particular, when restricted to the totally symmetric subspace, our bound reproduces the one developed for the post-selection technique \cite{Renner2008, Christandl2009}.
The second case we consider is when we perform independent and identical quantum measurements on an arbitrary quantum state prepared by an adversary.  In this case, we have an inequality between the probability mass function for an empirical probability of the measurement outcomes and a mixture of multinomial distribution up to a polynomial factor.  
We then obtain a Sanov-type concentration inequality for the probability of obtaining an empirical probability that lies in a set ${\cal A}$ of atypical events. 
\setcounter{theorem}{2}
\begin{theorem}[Informal]
    Let ${\cal M}$ be a $k$-outcome measurement channel and ${\cal P}$ be the probability $(k-1)$-simplex.  For a state $\rho$, ${\cal M}(\rho)$ corresponds to a point in ${\cal P}$.  Let ${\cal R}\subset {\cal P}$ be a region that is bounded by a hyperplane and contains $\{{\cal M}(\rho):\forall \rho\}$ and all the vertices of ${\cal P}$ except one.  Expand ${\cal R}$ by moving the boundary between ${\cal R}$ and ${\cal P}\setminus{\cal R}$ by ${\cal O}(n^{-1})$ to define the region ${\cal R}'$.  Let $P_{\bm{x}}$ denotes the empirical probability for a sequence $\bm{x}$.  Then, for any closed convex subset ${\cal A}\subseteq {\cal P}$ and for any $n$-partite quantum state $\rho^n$, the sequence $\bm{x}$ of outcomes of the measurement ${\cal M}^{\otimes n}$ performed on $\rho^n$ satisfies
    \begin{equation}
        {\cal M}^{\otimes n}(\rho^n)[P_{\bm{x}}\in{\cal A}] \leq \max_{\bm{p}\in{\cal A}}\max_{\bm{q}\in{\cal R}'} f_c(n,k)e^{-n D(\bm{p}\|\bm{q})},
    \end{equation}
    where $f_c(n,k)= {\cal O}(n^{(k-1)/2})$ for a large $n$ and $D(\cdot\|\cdot)$ denotes the Kullback-Leibler divergence.
\end{theorem}

\noindent As opposed to the first case, the polynomial factor in this case does not depend on the dimension of the underlying quantum system but depends on the number of measurement outcomes.

Both of our bounds are thus Sanov-type and thus expected to be tighter than those obtained by the Azuma-type inequalities.  To numerically demonstrate this, we consider two concrete scenarios.  The first one is a simple estimation task to obtain an upper bound on a random variable defined through the quantum measurements.  The second one is a finite-size key rate of a quantum key distribution protocol.  In both cases, we compare the performance obtained through conventional concentration inequalities and our new concentration inequalities.

Applicability of our results is not restricted to a finite-size analysis of QKD.  In the verifiable blind quantum computation~\cite{Takeuchi2018} and learning of a non-i.i.d.~quantum state~\cite{Fawzi2024} as mentioned earlier, the failure probability of a randomly-permuted property test on an adversarial quantum state is estimated.  Our results may fit to these cases as well, where one cares about exponentially rare outcomes of measurements performed on an adversarial state.

\setcounter{theorem}{0}
\subsection{Summary of our main results}\label{sec:summary_of_results}
We obtained three theorems (Theorems \ref{theo:refined_de_finetti}--\ref{theo:reduction_classical_iid}) for the concentration inequalities in the adversarial setups.
Theorems~\ref{theo:refined_de_finetti} and \ref{theo:de_finetti_with_symmetry} consider the case where the quantum state is symmetric under the permutation of subsystems, possibly with an additional symmetry restriction on each subsystem (Theorem~\ref{theo:de_finetti_with_symmetry}).  
These two theorems are given in terms of operator inequalities to show that a normalized density operator of any permutation-symmetric states is dominated, up to a polynomial factor in the number of subsystems, by a density operator for a mixture of i.i.d.~quantum states.  Such a theorem can immediately be combined with any large-deviation theorem for quantum measurements on i.i.d.~states to derive a bound for non-i.i.d.~but permutation-symmetric states.
The polynomial factor can be viewed as a price in extending the applicability to the non-i.i.d.~case.  Note that this scenario works even in the case where non-i.i.d.~quantum measurements are performed.
To the best of the authors' knowledge, this approach was first used in the context of quantum information theory for proving the security of the Bennett 1992 (B92) QKD protocol against general attacks \cite{Tamaki2003}.  A more general form was derived in relation to quantum extension \cite{Renner2007, Renner2008, Christandl2009, Fawzi2015} of the de Finetti theorem, which is given by \cite{Christandl2009}
\begin{align}
    \rho_{A^n} &\leq {n+d^2-1 \choose n} \int  \sigma_A^{\otimes n}\mu(\sigma), \label{eq:conventional}
\end{align}
where $\rho_{A^n}$ is invariant under the permutation of $A^n$, $d$ is the dimension of the system $A$, and $\mu$ denotes a probability measure on the set of density operators on the system $A$.
Compared to this known inequality, Theorem~\ref{theo:refined_de_finetti} provides an improvement in the polynomial factor as well as a further refinement of the factor by allowing its dependence on the i.i.d.~state, and applicability to a subsystem which makes it possible to compose multiple uses of the theorem when two or more subsystems separately adhere to permutation symmetry.  Theorem~\ref{theo:de_finetti_with_symmetry} is an adaptation of Theorem~\ref{theo:refined_de_finetti} to the more restricted cases where each subsystem adheres to a general symmetry in addition to the permutation symmetry among $n$ systems.  This is in contrast to the previous works~\cite{Leverrier2009, Leverrier2018, Gross2021} in which more general symmetries than the permutation among $n$ systems were considered.
Finally, our third result, Theorem~\ref{theo:reduction_classical_iid}, treats a somewhat different setup: a concentration inequality for the measurement outcomes when independent and identical measurements are performed on an adversarial quantum state.  Unlike the previous two theorems, this theorem does not require any symmetry on the quantum system as long as the concentration phenomenon for the empirical probability of measurement outcomes is considered.  Furthermore, the resulting statement does not depend on the dimension of the quantum system but depends on the number of measurement outcomes.

Our first result considers an operator inequality to bound a permutation-invariant state from above by a mixture of i.i.d.~quantum states, which is applicable under the same prerequisite as the conventional de-Finetti-type inequality~\eqref{eq:conventional}.
Let ${\cal H}_A$ be a $d$-dimensional Hilbert space.  Let $Y_n^d$ be the set of Young diagrams with $n$ boxes and at most $d$ rows, i.e.,
\begin{align}
    Y_n^d &\coloneqq \{\bm{n}=(n_1,\ldots,n_d)\in Z_n^d:n_1\geq n_2 \geq \cdots \geq n_d\}\label{eq:def_Y_n^d}
\end{align}
with 
\begin{equation}
    \begin{split}
    Z_n^d &\coloneqq \biggl\{\bm{n}=(n_1,\ldots,n_d)\in\{0,\ldots,n\}^{\times d}: \\
    & \hspace{4cm} \sum_{i=1}^{d}n_i=n\biggr\},
    \end{split}\label{eq:def_Z_n^k}
\end{equation}
where $i$-th element of $\bm{n}\in Y_n^d$ denotes the number of boxes in $i$-th row.
From the Schur-Weyl duality, ${\cal H}_A^{\otimes n}$ can be decomposed as
\begin{equation}
    {\cal H}_A^{\otimes n} = \bigoplus_{\bm{n}\in Y_n^d} {\cal U}_{\bm{n}} \otimes {\cal V}_{\bm{n}},  \label{eq:hilb_space_schur_weyl}
\end{equation}
where ${\cal U}_{\bm{n}}$ denotes an irreducible representation space of $\mathrm{U}(d)$ and ${\cal V}_{\bm{n}}$ denotes that of the permutation group $S_n$.  Let $P_{\bm{n}}$ be a projection onto ${\cal U}_{\bm{n}}\otimes{\cal V}_{\bm{n}}$.  It is known that a state that is invariant under the permutation of $n$ subsystems of $A^n$ has a direct-sum structure over $\bm{n}$ in Eq.~\eqref{eq:hilb_space_schur_weyl} and thus commutes with $P_{\bm{n}}$. 
This fact holds even when the permutation-invariant state on the system $A^n$ entangles with an environmental system $E$. 
That is, if we define $\rho_{A^n E}$ as a density operator on ${\cal H}_A^{\otimes n}\otimes {\cal H}_E$ invariant under the permutation of $n$ subsystems of $A^n$, then the following holds:
\begin{equation}
    \rho_{A^n E} = \sum_{\bm{n}\in Y_n^d} P_{\bm{n}}\rho_{A^n E}P_{\bm{n}},\label{eq:A_n_E_direct_sum}
\end{equation}
where the abbreviated notation $P_{\bm{n}}$ acting on ${\cal H}_A^{\otimes n}\otimes {\cal H}_E$ means that it acts trivially on ${\cal H}_E$.
From now on, if $X_{AB}$ is defined as an operator on ${\cal H}_A\otimes{\cal H}_B$, then $X_A$ denotes $\mathrm{Tr}_B[X_{AB}]$. 
A map $\Phi_A$ with the index $A$ denotes the completely positive trace-preserving (CPTP) map $\Phi$ from the set of operators on the system $A$ to itself. 
For a positive operator $\tau$, $\mathrm{supp}(\tau)$ denotes the support projection, i.e., the minimal projection $Q$ that satisfies $\mathrm{Tr}[Q\tau]=\mathrm{Tr}[\tau]$.
Let ${\rm Haar}(d)$ denote the Haar measure on the $d$-dimensional unitary group $\mathrm{U}(d)$.
For a $d$-tuple of real numbers $\bm{t}$, let $\diag(\bm{t})$ denote a density operator represented by a diagonal matrix with the diagonal entries $\bm{t}$ in a fixed basis of ${\cal H}_A$.
The following functions frequently appear in the later analyses.
\begin{definition}\label{def:functions_used}
    For $n, d\in\mathbb{N}$ and $\bm{n}\in Y_n^d$, let $s_{\bm{n}}(\bm{t})$ be the Schur function of $\bm{t}\in\mathbb{R}^{\times d}$ with the shape $\bm{n}$ defined as 
    \begin{equation}
        s_{\bm{n}}(\bm{t}) \coloneqq \frac{\det (t_i^{n_j + d - j})_{1\leq i,j \leq d}}{\det(t_i^{d - j})_{1\leq i,j \leq d}}.\label{eq:Schur_func}
    \end{equation}
    Then, the function $f_q(n,d;\bm{n})$ is defined with $s_{\bm{n}}(\bm{t})$ as 
    \begin{equation}
        f_q(n,d;\bm{n}) \coloneqq \frac{\dim{\cal U}_{\bm{n}}}{s_{\bm{n}}(\bm{n}/n)\dim{\cal V}_{\bm{n}}}.\label{eq:n-dep_coef_quantum}
    \end{equation}
    Furthermore, let $f_q(n,d)$ be defined as 
    \begin{align}
        f_q(n,d) \coloneqq \frac{(n+d-1)^{\frac{(d^2-1)}{2}}}{\sqrt{2\pi (d/e^2)^{d}}\, G(d+1)}, \label{eq:coef_quantum}
    \end{align}
    where $G(m)$ denotes the Barnes G-function given for an integer $m\geq 2$ by 
    \begin{equation}
        G(m)= \prod_{i=0}^{m-2} i!. \label{eq:Barnes_G}
    \end{equation}
    For $\bm{t}=(t_1,\ldots,t_d)\geq 0$ with $\sum_{i=1}^d t_i=1$ and $\bm{m}=(n_1,\ldots,n_d)\in Z_n^d$, let $\mathrm{Mult}_{\bm{t}}(\bm{m})$ be the probability density function of the multinomial distribution given by 
    \begin{equation}
        \mathrm{Mult}_{\bm{t}}(\bm{m})\coloneqq \frac{n!}{n_1!\cdots n_d!}\prod_{i=1}^d t_i^{n_i}. \label{eq:multinomial_pdf}
    \end{equation}
    Then, we define $f_c(n,d;\bm{m})$ as 
    \begin{equation}
        f_c(n,d;\bm{m})\coloneqq [\mathrm{Mult}_{\bm{m}/n}(\bm{m})]^{-1}, \label{eq:n-dep_coef_classical}
    \end{equation}
    and define $f_c(n,d)$ as
    \begin{equation}
        f_c(n,d) \coloneqq \frac{n^{\frac{d-1}{2}}}{\sqrt{2\pi (d/e^2)^{d}}}. \label{eq:coef_classical}
    \end{equation}
\end{definition}
The following relations between functions defined above will be proved later.
\begin{lemma} \label{lem:ineq_betwee_funcs}
    Let $f_q(n,d;\bm{n})$, $f_q(n,d)$, $f_c(n,d;\bm{m})$, and $f_c(n,d)$ be as defined in Definition~\ref{def:functions_used}.  Then, we have 
    \begin{equation}
        f_q(n,d;\bm{n}) \leq f_q(n,d) \label{eq:ineq_between_quantum_funcs}
    \end{equation}
    and 
    \begin{equation}
        f_c(n,d;\bm{m}) \leq f_c(n,d). \label{eq:ineq_between_classical_funcs}
    \end{equation}
\end{lemma}
Now, we state our first result.
\begin{theorem}\label{theo:refined_de_finetti}
    Consider a composite system $A^nE$ associated with a Hilbert space ${\cal H}_A^n\otimes {\cal H}_E$ with $\dim{\cal H}_A=d$. 
    Then, there exists a pair $({\cal S}_{A^n},{\cal R}_{A^n})$ of CPTP maps that satisfies the following conditions (i), (ii), and (iii).
    \medskip
    
        \noindent (i)~For any density operator $\rho_{A^nE}$ and for any $\bm{n}\in Y_n^d$, we have
        \begin{equation}
            {\cal S}_{A^n}\otimes \mathrm{Id}_E(P_{\bm{n}}\rho_{A^n E}P_{\bm{n}})=P_{\bm{n}}{\cal S}_{A^n}\otimes \mathrm{Id}_E(\rho_{A^n E})P_{\bm{n}}.\label{eq:commutativity_with_proj}
        \end{equation}
        
        \noindent (ii)~For any $\rho_{A^n E}$ that is invariant under permutation of $n$ systems of $A^n$, we have 
        \begin{align}
            {\cal R}_{A^n}\circ {\cal S}_{A^n}\otimes \mathrm{Id}_E(\rho_{A^n E})=\rho_{A^n E}.\label{eq:extension_condition} 
        \end{align}

        \noindent (ii)~For any $\rho_{A^n}$ that is invariant under permutation of $n$ systems of $A^n$ and for any $\bm{n}\in Y_n^d$, we have
        \begin{equation}
            \begin{split}
            &\mathrm{supp}({\cal S}_{A^n}(P_{\bm{n}}\rho_{A^n}P_{\bm{n}})) \\
            &\leq f_q(n,d;\bm{n})\\
            &\qquad \times{\cal S}_{A^n}\!\Bigl(\mathbb{E}_{U_A \sim {\rm Haar}(d)}\Bigl[\Bigl(U_A\,\diag\Bigl(\frac{\bm{n}}{n}\Bigr)\,U_A^{\dagger}\Bigr)^{\otimes n}\Bigr]\Bigr),\label{eq:support_upper_bound}
            \end{split}
        \end{equation} 
        where $f_q(n,d;\bm{n})$ is defined in Definition~\ref{def:functions_used}.
    
\end{theorem}

This theorem can immediately be cast into the conventional form of Eq.~\eqref{eq:conventional}.
Since ${\cal S}_{A^n}$ is a CPTP map, we trivially have, for any $\bm{n}\in Y_n^d$,
\begin{equation}
    {\cal S}_{A^n}(P_{\bm{n}}\rho_{A^n}P_{\bm{n}}) \leq \mathrm{Tr}[P_{\bm{n}}\rho_{A^n}]\; \mathrm{supp}({\cal S}_{A^n}(P_{\bm{n}}\rho_{A^n}P_{\bm{n}})),\label{eq:trivial_bound_by_supp}
\end{equation}
and thus, for any permutation-invariant state $\rho_{A^n}$, we have 
\begin{align}
    {\cal S}_{A^n}(\rho_{A^n}) &= \sum_{\bm{n}\in Y_n^d} {\cal S}_{A^n}(P_{\bm{n}}\rho_{A^n}P_{\bm{n}}) \\
    & \leq \sum_{\bm{n}\in Y_n^d}\mathrm{Tr}[P_{\bm{n}}\rho_{A^n}]\; \mathrm{supp}({\cal S}_{A^n}(P_{\bm{n}}\rho_{A^n}P_{\bm{n}})).\label{eq:rho_bounded_prob_supp}
\end{align}
Combining the above with Eqs.~\eqref{eq:extension_condition} and \eqref{eq:support_upper_bound}, we have
\begin{equation}
    \begin{split}
    \rho_{A^n} &\leq  \sum_{\bm{n}\in Y_n^d} f_q(n,d;\bm{n})\mathrm{Tr}[P_{\bm{n}}\rho_{A^n}]\\
    & \qquad \times \mathbb{E}_{U_A\sim {\rm Haar}(d)}\Bigl[\Bigl(U_A\,\diag\Bigl(\frac{\bm{n}}{n}\Bigr)\,U_A^{\dagger}\Bigr)^{\otimes n}\Bigr], 
    \end{split}\label{eq:refined_de_finetti}
\end{equation}
where we used the fact that $\tau_A^{\otimes n}$ is permutation invariant and thus satisfies ${\cal R}_{A^n}\circ{\cal S}_{A^n}(\tau_A^{\otimes n})=\tau_A^{\otimes n}$.
We see that Eq.~\eqref{eq:refined_de_finetti} takes a similar form to Eq.~\eqref{eq:conventional} but with different coefficients $f_q(n,d;\bm{n})$.  For comparison, let us first reformulate the right-hand side of Eq.~\eqref{eq:refined_de_finetti} using Lemma~\ref{lem:ineq_betwee_funcs} as
\begin{equation}
    \begin{split}
    \rho_{A^n} &\leq   f_q(n,d)\sum_{\bm{n}\in Y_n^d}\mathrm{Tr}[P_{\bm{n}}\rho_{A^n}]\\
    & \qquad \times \mathbb{E}_{U_A\sim {\rm Haar}(d)}\Bigl[\Bigl(U_A\,\diag\Bigl(\frac{\bm{n}}{n}\Bigr)\,U_A^{\dagger}\Bigr)^{\otimes n}\Bigr].
    \end{split}\label{eq:refined_de_finetti_mixture_iid}
\end{equation}
Then, the right-hand side is in the form of a probabilistic mixture of i.i.d.~states up to the factor $f_q(n,d)$.
When $n\gg d$, which is of our interest, the function $f_q(n,d)$ scales as $n^{(d^2-1)/2}$ whereas the coefficient in the conventional bound~\eqref{eq:conventional} scales as $n^{d^2-1}$.  Our bound thus achieves the square-root improvement in this case.

The $\bm{n}$-dependent coefficient of the bound in Eq.~\eqref{eq:refined_de_finetti} may also be beneficial when there is a restriction on the probability mass function $\mathrm{Tr}[P_{\bm{n}}\rho_{A^n}]$ over $\bm{n}\in Y_n^d$ or when one performs a quantum measurement on the state that may depend on the spectrum of the i.i.d.~state.  For example, consider the case $\mathrm{Tr}[P_{\bm{n}}\rho_{A^n}]=\delta_{\bm{n}\bm{n}_0}$ with $\bm{n}_0=(n,0,\ldots,0)$, which amounts to restricting $\rho_{A^n}$ supported only on the symmetric subspace $\mathrm{Sym}^n({\cal H}_A)$ of ${\cal H}_A^{\otimes n}$.  In this case, we have $\dim {\cal U}_{\bm{n}_0}={n+d-1\choose n}$, $\dim {\cal V}_{\bm{n}_0}=1$, and $s_{\bm{n}_0}(\bm{n}_0/n)=1$, and the density operator $\diag(\bm{n}_0/\bm{n})$ corresponds to a pure state.  From Eq.~\eqref{eq:refined_de_finetti} and Lemma~\ref{lem:ineq_betwee_funcs}, we then have 
\begin{equation}
    \rho_{A^n} \leq {n+d-1 \choose n} \int d\phi \ket{\phi}\!\bra{\phi}_{A}^{\otimes n},\label{eq:symmetric_subspace}
\end{equation}
where $d\phi$ denotes the Haar measure on the set of pure states in the system $A$.
This inequality is exactly what was proved in Ref.~\cite{Christandl2009}.  Combining the above with the fact that any permutation-invariant state in the system $A^n$ has a purification in the symmetric subspace $\mathrm{Sym}^n({\cal H}_A\otimes{\cal H}_B)$ with ${\cal H}_B\cong{\cal H}_A$ \cite{Renner2008, Christandl2009}, they derived Eq.~\eqref{eq:conventional}.  Our Theorem~\ref{theo:refined_de_finetti} can thus be regarded as a generalization of the operator inequality obtained in Ref.~\cite{Christandl2009}.


The reason why we used the support projection in Theorem~\ref{theo:refined_de_finetti} is to make it composable when two or more parts of the system undergo separate permutations.  Consider a system $A^n B^m$ and a state $\rho_{A^nB^m}$ that is invariant under the permutations of $n$ systems in $A^n$ and also is invariant under the permutations of $m$ systems in $B^m$.  In such a case, a pair of inequalities like $\rho_{A^nB^m}\leq X_{A^n}\otimes I_{B^m}$ and $\rho_{A^nB^m} \leq I_{A^n}\otimes X_{B^m}$ would not imply $\rho_{A^nB^m} \leq X_{A^n}\otimes X_{B^m}$.  On the other hand, in the case of support projections, it holds that $\mathrm{supp}(\rho_{A^nB^m})\leq \mathrm{supp}(\rho_{A^n})\otimes \mathrm{supp}(\rho_{B^m})$.  Thanks to this property, we can apply Theorem~\ref{theo:refined_de_finetti} with $(A,n,E)\Rightarrow (A,n,B^n)$ and $(A,n,E)\Rightarrow(B,m,A^n)$ to obtain
\begin{align}
    {\cal R}_{A^n}\circ{\cal S}_{A^n} \otimes {\cal R}_{B^m}\circ {\cal S}_{B^m}(\rho_{A^nB^m}) = \rho_{A^nB^m},\label{eq:composable_identity}
\end{align}
and for any $\bm{n}\in Y_n^d$ and $\bm{m}\in Y_m^{d'}$, we have 
\begin{equation}
    \begin{split}
    &{\cal S}_{A^n} \otimes {\cal S}_{B^m}(P_{\bm{n}}\otimes P_{\bm{m}}\rho_{A^nB^m}P_{\bm{n}}\otimes P_{\bm{m}})\\
    & = P_{\bm{n}}\otimes P_{\bm{m}}{\cal S}_{A^n} \otimes {\cal S}_{B^m}(\rho_{A^nB^m})P_{\bm{n}}\otimes P_{\bm{m}} \label{eq:composite_commutativity}
    \end{split}
\end{equation}
and
\begin{align}
    &\mathrm{supp}\bigl({\cal S}_{A^n}\otimes {\cal S}_{B^m}(P_{\bm{n}}\otimes P_{\bm{m}}\rho_{A^nB^m}P_{\bm{n}}\otimes P_{\bm{m}})\bigr) \nonumber \\
    &\leq \mathrm{supp}({\cal S}_{A^n}(P_{\bm{n}}\rho_{A^n}P_{\bm{n}})) \otimes \mathrm{supp}({\cal S}_{B^m}(P_{\bm{m}}\rho_{B^m}P_{\bm{m}})) \\
    \begin{split}
    & \leq f_q(n,d;\bm{n})f_q(m,d';\bm{m}) \\
    &\quad \times {\cal S}_{A^n}\!\biggl(\mathbb{E}_{U_A\sim {\rm Haar}(d)}\Bigl[\Bigl(U_A\,\diag\Bigl(\frac{\bm{n}}{n}\Bigr)\,U_A^{\dagger}\Bigr)^{\otimes n}\Bigr]\biggr)   \\
    &\qquad \otimes {\cal S}_{B^m}\!\biggl(\mathbb{E}_{U_B\sim {\rm Haar}(d')}\Bigl[\Bigl(U_B\,\diag\Bigl(\frac{\bm{m}}{m}\Bigr)\,U_B^{\dagger}\Bigr)^{\otimes m}\Bigr]\biggr), \label{eq:composite_support_bound}
    \end{split}
\end{align}
from which we can derive an inequality of the form $\rho_{A^nB^n}\leq X_{A^n}\otimes X_{B^n}$ as we obtained Eq.~\eqref{eq:refined_de_finetti}.

Our next result is when a local system $A$ has a (unitary) symmetry restriction.  Suppose that a group $G$ has a unitary action on ${\cal H}_A$ with $\dim {\cal H}_A=d$, namely, we have a group representation $\pi:G\to B({\cal H}_A)$.  Since any unitary representation of a group is completely decomposable, $\pi$ is a direct sum of irreducible subrepresentations.  Labeling the inequivalent representations appearing in the decomposition by integers $\{1,\ldots,k\}$, the space ${\cal H}_A$ can be decomposed as
\begin{equation}
    {\cal H}_A = \bigoplus_{j=1}^{k} {\cal H}_j^{R}\otimes {\cal H}_j^{D},\label{eq:decomp_local_system_relabel}
\end{equation}
under which the representation $\pi$ is decomposed as
\begin{equation}
    \pi(g)=\bigoplus_{j=1}^k \pi_j(g)\otimes I_j^D, \label{eq:rep_of_local_symmetry}
\end{equation}
where $\pi_j:G\to B(H_j^R)$ is an irreducible representation with dimension $r_j=\dim{\cal H}_j^R$ and has degeneracy $d_j=\dim {\cal H}_j^D$ in $\pi$.  These numbers are related to the total dimension $d$ by 
\begin{equation}
    d=\sum_{j=1}^{k} r_j d_j.
\end{equation}
We consider the case where each of the $n$ systems in $A^n$ is independently subject to the symmetry restriction associated with $G$.  The symmetry of the whole $n$ systems is then associated with the direct product $G^{\times n}$ and the tensor product representation $\pi^{\otimes n}$ of it.  We say an operator $X_{A^n E}$ acting on ${\cal H}_{A^n}\otimes {\cal H}_E$ is locally $G$-invariant when $\pi^{\otimes n}(\bm{g})\,X_{A^n E}\,\pi^{\otimes n}(\bm{g})^{\dagger}=X_{A^n E}$ holds for all $\bm{g}\in G^{\times n}$.

By decomposing the space of each subsystem in $A^n$ via Eq.~\eqref{eq:decomp_local_system_relabel}, the space ${\cal H}_A^{\otimes n}$ is decomposed into a sum over an $n$-tuple of integers $\bm{j}=(j_1,\ldots,j_n)$ as 
\begin{equation}
    {\cal H}_A^{\otimes n}=\bigoplus_{\bm{j}\in\{1,\ldots,k\}^{\times n}} {\cal H}_{\bm{j}}^R \otimes {\cal H}_{\bm{j}}^D \label{eq:decomp_H_bm_j}
\end{equation}
with
\begin{equation}
    {\cal H}_{\bm{j}}^{R(D)} \coloneqq \bigotimes_{i=1}^{n} {\cal H}_{j_i}^{R(D)}. \label{eq:def_H_bm_j}
\end{equation}
The unitary $\pi^{\otimes n}(\bm{g})$ with $\bm{g}\in G^{\times n}$ acts on this space as 
\begin{equation}
    \pi^{\otimes n}(\bm{g}) = \bigoplus_{\bm{j}\in\{1,\ldots,k\}^{\times n}} \pi_{\bm{j}}(\bm{g})\otimes I_{{\cal H}_{\bm{j}}^D},
\end{equation}
where $\pi_{\bm{j}}(\bm{g})=\bigotimes_{i=1}^n \pi_{j_i}(g_i)$ is an irreducible representation of $G^{\times n}$.
We group the summation according to the frequency of each label $l\in\{1,\ldots,k\}$ in the $n$-tuple $\bm{j}$, or the type of $\bm{j}$.  
For $Z_n^k$ defined in Eq.~\eqref{eq:def_Z_n^k}, we define a map $\bm{\chi}$ as
\begin{equation}
    \begin{split}
        \bm{\chi}:\{1,\ldots,k\}^{\times n}&\to Z_n^k \\
        \bm{j} &\mapsto (|\{i\in\{1,\ldots,n\}: j_i=1\}|, \\
        &\qquad \ldots,|\{i\in\{1,\ldots,n\}:j_i=k\}|)
    \end{split} \label{eq:def_bm_chi}
\end{equation}
and denote its preimage by 
\begin{equation}
    {\cal T}(\bm{m})\coloneqq \bm{\chi}^{-1}(\bm{m}) \label{eq:def_type_set}
\end{equation}
for $\bm{m}\in Z_n^k$.  The cardinality $|{\cal T}(\bm{m})|$ of ${\cal T}(\bm{m})$ for $\bm{m}=(n_1,\ldots,n_k)$ is given by
\begin{equation}
    |{\cal T}(\bm{m})| = \frac{n!}{n_1!\cdots n_k!}. \label{eq:cardinality_type}
\end{equation}
Note that these sets give a grouping of the $n$-tuple $\bm{j}$, namely, $\{1,\ldots,k\}^{\times n}=\bigsqcup_{\bm{m}\in Z_n^k} {\cal T}(\bm{m})$.  Now, the space ${\cal H}_A^{\otimes n}$ is decomposed as 
\begin{equation}
    {\cal H}_A^{\otimes n} = \bigoplus_{\bm{m}\in Z_n^k} \bigoplus_{\bm{j}\in {\cal T}(\bm{m})} {\cal H}_{\bm{j}}^R \otimes {\cal H}_{\bm{j}}^D.
\end{equation}
For $\bm{m}=(n_1,\ldots,n_k)\in Z_n^k$, let us choose a representative $\bm{j}_0(\bm{m})$ of the set ${\cal T}(\bm{m})$ as
\begin{equation}
    \bm{j}_0(\bm{m}) \coloneqq (\underbrace{1,\ldots,1}_{n_1},\underbrace{2,\ldots,2}_{n_2},\ldots,\underbrace{k,\ldots,k}_{n_k}).\label{eq:def_bm_j_0}
\end{equation}
For this particular sequence, the corresponding Hilbert space can simply be written as
\begin{align}
    &{\cal H}_{\bm{j}_0(\bm{m})}^R\otimes {\cal H}_{\bm{j}_0(\bm{m})}^D \nonumber \\
    &= \bigotimes_{j=1}^k \bigl({\cal H}_j^R \otimes {\cal H}_j^D\bigr)^{\otimes n_j} \\
    &= \bigotimes_{j=1}^k \bigl({\cal H}_{j}^R \bigr)^{\otimes n_j} \otimes \Biggl(\bigoplus_{\bm{n}_j\in Y_{n_j}^{d_j}}{\cal U}_{\bm{n}_j}\otimes {\cal V}_{\bm{n}_j}\Biggr), \label{eq:decomp_for_j_0}
\end{align}
where we applied the decomposition of Eq.~\eqref{eq:hilb_space_schur_weyl} to $({\cal H}_j^D)^{\otimes n_j}$.  Other members of ${\cal T}(\bm{m})$ are connected to $\bm{j}_0(\bm{m})$ by permutations.  That is to say, we may choose a permutation $\sigma_{\bm{j}}$ for each $\bm{j}\in \{1,\ldots,k\}^{\times n}$ to satisfy 
\begin{align}
    \sigma_{\bm{j}}(\bm{j}_0(\bm{\chi}(\bm{j}))) =\bm{j}. \label{eq:def_sigma_bm_j}
\end{align}
Such a choice is not unique in general, but the following argument holds independent of the choice.
Let $V_{\sigma_{\bm{j}}}=V_{\sigma_{\bm{j}}}^R \otimes V_{\sigma_{\bm{j}}}^D$ be a unitary representation of the permutation $\sigma_{\bm{j}}$, where $V_{\sigma_{\bm{j}}}^{R(D)}$ acts on the Hilbert space $\bigl(\bigoplus_{j=1}^k{\cal H}_j^{R(D)}\bigr)^{\otimes n}$.  Then, we have  
\begin{equation}
    {\cal H}_{\bm{j}}^{R(D)} = V_{\sigma_{\bm{j}}}^{R(D)} {\cal H}_{\bm{j}_0(\bm{\chi}(\bm{j}))}^{R(D)}. \label{eq:unitary_connect_j_j_0}
\end{equation}
We note here that the Hilbert space $\bigl(\bigoplus_{j=1}^k{\cal H}_j^{R}\bigr)^{\otimes n}\otimes \bigl(\bigoplus_{j=1}^k{\cal H}_j^{D}\bigr)^{\otimes n}$ on which the action of $V_{\sigma_{\bm{j}}}$ is defined is larger than the original Hilbert space ${\cal H}_A^{\otimes n}$.  However, it keeps the subspace ${\cal H}_A^{\otimes n}$ invariant, and thus we can consider the restriction of $V_{\sigma_{\bm{j}}}$ onto ${\cal H}_A^{\otimes n}$.
Combining Eqs.~\eqref{eq:decomp_H_bm_j}, \eqref{eq:decomp_for_j_0}, and \eqref{eq:unitary_connect_j_j_0}, we arrive at the decomposition associated with the permutation group and the group $G^{\times n}$:
\begin{equation}
    {\cal H}_A^{\otimes n} = \bigoplus_{\substack{\bm{m}=(n_1,\ldots,n_k)\\ \in Z_n^k}} \bigoplus_{\substack{(\bm{n}_1,\ldots,\bm{n}_k) \\ \in Y_{n_1}^{d_1}\cdots\times Y_{n_k}^{d_k}}} {\cal H}_{\bm{m} ;\bm{n}_1,\ldots,\bm{n}_k}, \label{eq:decomp_type_seq_irrep}
\end{equation}
with 
\begin{equation}
    {\cal H}_{\bm{m};\bm{n}_1,\ldots,\bm{n}_k}\coloneqq \sum_{\bm{j}\in{\cal T}(\bm{m})}\!V_{\sigma_{\bm{j}}}\!\left[ \bigotimes_{j=1}^k \bigl(H_{j}^R \bigr)^{\otimes n_j} \otimes \bigl({\cal U}_{\bm{n}_j}\otimes {\cal V}_{\bm{n}_j}\bigr)\right].
\end{equation}
Let $P_{\bm{m};\bm{n}_1,\ldots,\bm{n}_k}$ be the projection onto ${\cal H}_{\bm{m};\bm{n}_1,\ldots,\bm{n}_k}$.  
Our next theorem can then be stated as follows.

\begin{theorem}\label{theo:de_finetti_with_symmetry}
    Let $G$ be a group that has a unitary action on a Hilbert space ${\cal H}_A$ with $\dim{\cal H}_A=d$. 
   Then, under the decomposition of Eq.~\eqref{eq:decomp_type_seq_irrep}, any density operator $\rho_{A^n E}^{G}$ that is invariant under permutation of $n$ local systems of $A^n$ and also locally $G$-invariant satisfies
    \begin{align}
        \rho_{A^n E}^{G} = \sum_{\bm{m}\in Z_n^k}\sum_{\substack{(\bm{n}_1,\ldots,\bm{n}_k) \\ \in Y_{n_1}^{d_1}\cdots\times Y_{n_k}^{d_k}}} \rho^G_{A^n E}(\bm{m};\bm{n}_1,\ldots,\bm{n}_k), \label{eq:symmetric_direct_sum_decomp}
    \end{align}
    where
    \begin{align}
        \rho^G_{A^n E}(\bm{m};\bm{n}_1,\ldots,\bm{n}_k)\coloneqq  P_{\bm{m};\bm{n}_1,\ldots,\bm{n}_k} \rho_{A^n E}^{G} P_{\bm{m};\bm{n}_1,\ldots,\bm{n}_k}.  \label{eq:symmetric_direct_sum_decomp_2}
    \end{align}
    Furthermore, there exists a pair $({\cal S}_{A^n}^G, {\cal R}_{A^n}^G)$ of CPTP maps that satisfies the following conditions (i), (ii), and (iii).
    \medskip
    
    \noindent (i)~For any density operator $\rho_{A^nE}$ and for any $\bm{m}\in Z_n^k$ and $(\bm{n}_1,\ldots,\bm{n}_k)\in Y_{n_1}^{d_1}\times \cdots Y_{n_k}^{d_k}$, we have
    \begin{equation}
        \begin{split}
        &{\cal S}_{A^n}^G\otimes \mathrm{Id}_E(P_{\bm{m};\bm{n}_1,\ldots,\bm{n}_k}\rho_{A^n E}P_{\bm{m};\bm{n}_1,\ldots,\bm{n}_k}) \\
        &= P_{\bm{m};\bm{n}_1,\ldots,\bm{n}_k} {\cal S}_{A^n}^G\otimes \mathrm{Id}_E(\rho_{A^n E}) P_{\bm{m};\bm{n}_1,\ldots,\bm{n}_k}
        \end{split}\label{eq:symmetric_commutative_proj}
    \end{equation}

    \noindent (ii)~For any $\rho_{A^n E}^{G}$ that is invariant under permutation of $n$ local systems of $A^n$ as well as locally $G$-invariant, we have
    \begin{equation}
        {\cal R}_{A^n}^G \circ {\cal S}_{A^n}^G \otimes \mathrm{Id}_E (\rho_{A^n E}^G) = \rho_{A^n E}^G.\label{eq:recovery_symmetric}
    \end{equation}
    
    \noindent (iii)~For any $\rho_{A^n}^{G} $ that is invariant under permutation of $n$ local systems of $A^n$ as well as locally $G$-invariant and for any $\bm{m}\in Z_n^k$ 
    and $(\bm{n}_1,\ldots,\bm{n}_k)\in Y_{n_1}^{d_1}\times \cdots Y_{n_k}^{d_k}$, we have
    \begin{align}
        \begin{split}
        &\mathrm{supp}\Bigl({\cal S}_{A^n}^G \bigl(\rho^G_{A^n}(\bm{m};\bm{n}_1,\ldots,\bm{n}_k)\bigr)\Bigr)  \\
        &\leq f_c(n,k;\bm{m})\prod_{j=1}^k f_q(n_j,d_j;\bm{n}_j)\\
        & \qquad \times {\cal S}_{A^n}^G\bigl(\mathbb{E}_{\bm{U}\sim \bigtimes_{j=1}^k \mathrm{Haar}(d_j)}[\tau_{\bm{U}}(\bm{n}_1,\ldots,\bm{n}_k)^{\otimes n}]\bigr),
        \end{split}\label{eq:support_bound_symmetric}
    \end{align}
    where the density operator $\tau_{\bm{U}}(\bm{n}_1,\ldots,\bm{n}_k)$ is defined as
    \begin{align}
        \tau_{\bm{U}}(\bm{n}_1,\ldots,\bm{n}_k) &\coloneqq \bigoplus_{j=1}^k \frac{I_{{\cal H}_j^R}}{r_j} \otimes \Bigl(U_j\,\diag\Bigl(\frac{\bm{n}_j}{n}\Bigr)\,U_j^{\dagger}\Bigr)_{{\cal H}_j^{D}}. \label{eq:def_tau_U}
    \end{align}
\end{theorem}

Again, as a direct consequence of the theorem above with the fact that
\begin{align}
    &{\cal S}_{A^n}^G(\rho_{A^n}^G) \nonumber\\ &= \sum_{\bm{m}\in Z_n^k} \sum_{\substack{(\bm{n}_1,\ldots,\bm{n}_k)\\ \in Y_{n_1}^{d_1} \times\cdots\times Y_{n_k}^{d_k}}} {\cal S}_{A^n}^G \bigl(\rho_{A^n}^G(\bm{m};\bm{n}_1,\ldots,\bm{n}_k) \bigr) \\
    \begin{split}
    & \leq \sum_{\bm{m}\in Z_n^k}\sum_{\substack{(\bm{n}_1,\ldots,\bm{n}_k)\\ \in Y_{n_1}^{d_1} \times\cdots\times Y_{n_k}^{d_k}}}\mathrm{Tr}[\rho_{A^n}^G(\bm{m};\bm{n}_1,\ldots,\bm{n}_k)]\\
    &\qquad \times \mathrm{supp}\Bigl({\cal S}_{A^n}^G\bigl(\rho_{A^n}^G(\bm{m};\bm{n}_1,\ldots,\bm{n}_k)\bigr)\Bigr)
    \end{split}
\end{align}
always holds, we have
\begin{equation}
    \begin{split}
    \rho_{A^n}^G &\leq \sum_{\bm{m}\in Z_n^k} \sum_{\substack{(\bm{n}_1,\ldots,\bm{n}_k) \\ \in Y_{n_1}^{d_1} \times\cdots\times Y_{n_k}^{d_k}}} f_c(n,k;\bm{m}) \\ 
    &\qquad \prod_{j=1}^k f_q(n_j,d_j;\bm{n}_j)\mathrm{Tr}\!\left[P_{\bm{m};\bm{n}_1,\ldots,\bm{n}_k}\rho_{A^n}^G\right] \\
    &\qquad \quad \times \mathbb{E}_{\bm{U}\sim \bigtimes_{j=1}^k \mathrm{Haar}(d_j)}[\tau_{\bm{U}}(\bm{n}_1,\ldots,\bm{n}_k)^{\otimes n}].
    \end{split} \label{eq:concentration_with_symmetry}
\end{equation}
One may apply Lemma~\ref{lem:ineq_betwee_funcs} to obtain an elementary expression for the prefactor.

These two theorems enable us to reduce the problems of concentration inequalities on the permutation-invariant states to those of i.i.d.~states up to a polynomial factor.  This factor can be negligible for large $n$ since a concentration inequality typically yields exponential decay over $n$.  Concentration inequalities on the i.i.d.~state are relatively simple and tight bounds are known for these cases, which is beneficial for the finite-size analysis of the permutation-invariant states.

Our final result considers a different scenario from the previous two results.
Let us assume that an adversary prepares the quantum state $\rho^n$ in the $n$-composite system.
We perform independent and identical measurements ${\cal M}^{\otimes n}$ on the state.
The following theorem holds for the (classical) probability mass function of the measurement outcomes.

\begin{theorem}\label{theo:reduction_classical_iid}
    Consider a $k$-outcome measurement on a quantum system associated with a Hilbert space ${\cal H}$, specified by the POVM $\{M(i)\}_{i\in{\cal X}}$ with ${\cal X}\coloneqq \{1,\ldots,k\}$.
    Let ${\cal D}({\cal H})$ be the set of all the density operators on the Hilbert space ${\cal H}$.
    For $\rho\in{\cal D}({\cal H})$, let ${\cal M}(\rho): i \mapsto \mathrm{Tr}[M(i)\rho]$ be the probability mass function for the outcome of the measurement.  
    Consider a probability simplex ${\cal P}\subseteq \mathbb{R}^k$
    and define the region ${\cal Q} \subseteq {\cal P}$ as
    \begin{equation}
        \begin{split}
        {\cal Q}&\coloneqq \{(p_1,\ldots,p_k)\in{\cal P}: \\
        &\qquad \qquad \exists\rho\in{\cal D}({\cal H}) \text{ s.t.\ }\forall i \in{\cal X}, p_i={\cal M}(\rho)(i)\}.
        \end{split}\label{eq:possible_meas_outcomes}
    \end{equation}
    Assume $\bm{p}_1 \coloneqq (1,0,\ldots,0)\notin{\cal Q}$.
    Let ${\cal R}\subseteq {\cal P}$ be a region defined as
    \begin{equation}
        {\cal R}\coloneqq \{\bm{p}\in{\cal P}:\bm{u}\cdot\bm{p} \leq 0\}, \label{eq:def_curly_R}
    \end{equation}
    with $\bm{u}\coloneqq (1,-b_2,\ldots,-b_k)$, where $b_2,\ldots,b_k$ are nonnegative numbers chosen to satisfy ${\cal Q}\subseteq{\cal R}$.
    With $\bm{v}\coloneqq (1,1,\ldots,1)/\sqrt{k}$, define ${\cal R}'$ as 
    \begin{equation}
        {\cal R}' \coloneqq  \left\{\bm{p}\in{\cal P}:\bm{u}\cdot\bm{p} \leq \frac{\sqrt{2}\|\bm{u}-(\bm{v}\cdot\bm{u})\bm{v}\|}{n}\right\},\label{eq:region_sifted_by_N_inverse}
    \end{equation}
    where $\|\bm{a}\|$ denotes the Euclidean norm of the vector $\bm{a}$.

    Consider the measurement ${\cal M}^{\otimes n}$ on $n$ systems $A^n$, which has the POVM $\{M^{\otimes n}(\bm{i})\}_{\bm{i}\in{\cal X}^{\times n}}$ with $M^{\otimes n}\bigl((i_1,\ldots,i_n)\bigr)=M(i_1)\otimes\cdots\otimes M(i_n)$.  Let ${\cal A}\subseteq{\cal P}$ be a non-empty closed convex region of the probability simplex.
    Then, for any quantum state $\rho\in{\cal D}({\cal H}^{\otimes n})$, the probability mass function ${\cal M}^{\otimes n}(\rho):\bm{i}\mapsto \mathrm{Tr}[M^{\otimes n}(\bm{i})\rho]$ satisfies
    \begin{equation}
        \begin{split}
            &\sum_{\bm{i}:\bm{\chi}(\bm{i})/n\in{\cal A}}{\cal M}^{\otimes n}(\rho)(\bm{i}) \\
            & \leq \max_{\bm{p}\in{\cal A}, \bm{q}\in{\cal R}'} f_c(n,k)\exp[-n D(\bm{p}\|\bm{q})],
        \end{split}
    \end{equation}
    where $D(\cdot\|\cdot)$ denotes the Kullback-Leibler divergence, and the functions $\bm{\chi}(\bm{i})$ and $f_c(n,k)$ are defined respectively in Eqs.~\eqref{eq:def_bm_chi} and \eqref{eq:coef_classical}.
\end{theorem}

\begin{figure}[t]
    \centering
    \includegraphics[width=0.99\linewidth]{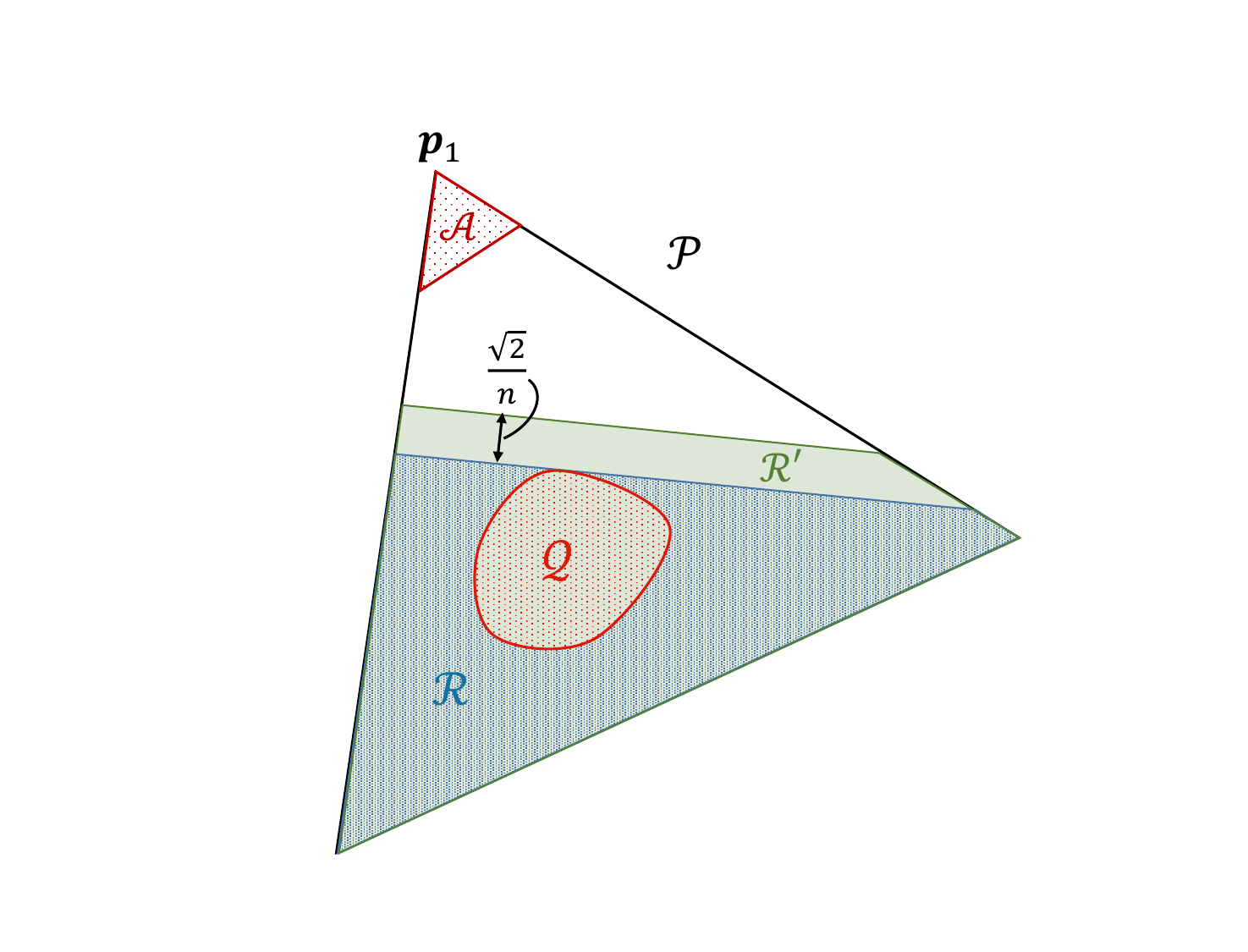}
    \caption{Geometrical relations between the (convex) subsets ${\cal Q}$, ${\cal R}$, ${\cal R}'$, and ${\cal A}$ of ${\cal P}$.  The subset ${\cal A}$ is non-empty, closed, and convex.  The subset ${\cal R}$ is bounded by a hyperplane and contains ${\cal Q}$ and all the extremal points of ${\cal P}$ except $\bm{p}_1=(1,0,\ldots,0)$.  The set ${\cal R}'$ consists of every point whose distance to ${\cal R}$ is no larger than $\sqrt{2}/n$.}
    \label{fig:simplex}
\end{figure}

Geometrical relations between the sets ${\cal Q}$, ${\cal R}$, ${\cal R}'$, and ${\cal A}$ are illustrated in Fig.~\ref{fig:simplex}.
This theorem gives a Sanov-type bound \cite{Sanov1957, Csiszar1984} even though the quantum state is prepared by an adversary.  Since Sanov-type inequalities are in general tighter than Azuma-Hoeffding-type inequalities \cite{Azuma_ineq, Kato2020}, which can also be applied in this setup, it may be useful in, for example, finite-size analyses of quantum cryptography.  
Note that Theorem~\ref{theo:reduction_classical_iid} also holds for a classical probability mass function ${\cal M}^n$ instead of ${\cal M}^{\otimes n}(\rho)$ as long as all the conditional probability mass function of ${\cal M}^n$ conditioned on $n-1$ outcomes are contained in a set ${\cal Q}\subseteq {\cal P}$, which replaces the precondition Eq.~\eqref{eq:possible_meas_outcomes} in this case.

\subsection{Organization of the paper}
In Sec.~\ref{sec:proofs_of_theorems}, we prove Theorems~\ref{theo:refined_de_finetti}--\ref{theo:reduction_classical_iid} stated above, where each subsection is assigned for each theorem.  In Sec.~\ref{sec:numerical_comparison}, we apply our results to simple examples and numerically demonstrate the tightness of our bounds by comparison with conventionally used bounds.  The first example in Sec.~\ref{sec:toy_model} is an estimation task of measurement outcomes with two non-orthogonal projections, which is not directly related to real applications but is still informative.  The second example is an application to the quantum key distribution protocol.  Finally, in Sec.~\ref{sec:discussion}, we summarize our results and discuss possible applications.

\section{Proofs of the main theorems}\label{sec:proofs_of_theorems}
\subsection{A permutation-invariant quantum state is bounded by a mixture of i.i.d.~quantum states}
In this section, we prove Lemma~\ref{lem:ineq_betwee_funcs} and Theorem~\ref{theo:refined_de_finetti}, which constructs an upper bound on a density operator of a permutation-invariant state with density operators of i.i.d.~states up to a multiplication factor. 
This has been intensively studied in the context of the quantum de Finetti theorem (see e.g.~\cite{Renner2007} for a review) under the name of post-selection technique \cite{Renner2008, Christandl2009, Fawzi2015}.  Ours can be regarded as a refinement and a generalization of the bound.

As defined prior to Theorem~\ref{theo:refined_de_finetti}, let ${\cal H}_A$ be a $d$-dimensional Hilbert space, $Y_n^d$ be the set of Young diagrams with $n$ boxes and at most $d$ rows whose elements can be labeled by $\bm{n}=(n_1, n_2, \ldots, n_d)$ with $n_1 \geq n_2 \geq \cdots n_d$ and $\sum_{i=1}^{d} n_i = n$.
As shown in Eq.~\eqref{eq:hilb_space_schur_weyl}, ${\cal H}_A^{\otimes n}$ can be decomposed into the direct sum of the subspaces in the form ${\cal U}_{\bm{n}} \otimes {\cal V}_{\bm{n}}$ labeled by $\bm{n}\in Y_n^d$.  Let $P_{\bm{n}}$ be a projection operator onto this subspace.
Let $\rho_{A^n E}$ be a state on ${\cal H}_A^{\otimes n}\otimes {\cal H}_E$ that is invariant under the permutation of $n$ local systems of $A^n$.  Then, from Schur's lemma, $\rho_{A^n E}$ has the form
\begin{equation}
    \rho_{A^n E} = \bigoplus_{\bm{n}\in Y_n^d}\mathrm{Tr}[P_{\bm{n}}\rho_{A^n E}]\,\rho_{{\cal U}_{\bm{n}} E} \otimes \frac{I_{{\cal V}_{\bm{n}}}}{\dim {\cal V}_{\bm{n}}},\label{eq:direct_sum_decomp}
\end{equation}
where $\rho_{{\cal U}_{\bm{n}}E}$ denotes a density operator on the subspace ${\cal U}_{\bm{n}}$ and $E$, and $I_{{\cal V}_{\bm{n}}}$ is an identity operator on the subspace ${\cal V}_{\bm{n}}$.  From this, it is obvious that the state $\rho_{A^n E}$ satisfies Eq.~\eqref{eq:A_n_E_direct_sum}.
We define a CPTP map ${\cal S}_{A^n}$ as follows:
\begin{align}
    \begin{split}
    {\cal S}_{A^n}(\tau_{A^n}) = \bigoplus_{\bm{n}\in Y_n^d} \mathrm{Tr}^{(\bm{n})}_{{\cal V}_{\bm{n}}} (P_{\bm{n}}\tau_{A^n}P_{\bm{n}}|_{{\cal U}_{\bm{n}}\otimes {\cal V}_{\bm{n}}}) \otimes \ket{0}\!\bra{0}_{{\cal V}_{\bm{n}}},
    \end{split}\label{eq:def_S}
\end{align}
where $\mathrm{Tr}^{(\bm{n})}$ denotes the trace acting on a matrix on ${\cal U}_{\bm{n}}\otimes {\cal V}_{\bm{n}}$, $\mathrm{Tr}^{(\bm{n})}_{{\cal V}_{\bm{n}}}$ denotes the partial trace over the space ${\cal V}_{\bm{n}}$, and $\ket{0}\!\bra{0}_{{\cal V}_{\bm{n}}}$ is an arbitrary pure state on ${\cal V}_{\bm{n}}$.  
We also define another CPTP map ${\cal R}_{A^n}$ as follows:
\begin{align}
    \begin{split}
    {\cal R}_{A^n}(\tau_{A^n}) = \bigoplus_{\bm{n}\in Y_n^d} \mathrm{Tr}^{(\bm{n})}_{{\cal V}_{\bm{n}}} (P_{\bm{n}}\tau_{A^n}P_{\bm{n}}|_{{\cal U}_{\bm{n}}\otimes {\cal V}_{\bm{n}}}) \otimes \frac{I_{{\cal V}_{\bm{n}}}}{\dim {\cal V}_{\bm{n}}}. 
    \end{split}\label{eq:def_R}
\end{align}

Now, let $\diag(\bm{t})$ be a density operator represented by a diagonal matrix with the spectrum $\bm{t}= (t_1,\ldots,t_d)$ in a fixed basis of ${\cal H}_A$.  Without loss of generality, we assume $t_1\geq \cdots \geq t_d (\geq 0)$.
Then, it is known that \cite{Fulton1991, Fulton1996, Hayashi2017} 
\begin{equation}
    \mathrm{Tr}[P_{\bm{n}}\diag(\bm{t})^{\otimes n}]=\dim {\cal V}_{\bm{n}}\; s_{\bm{n}}(\bm{t}), \label{eq:inf_coef}
\end{equation}
where $s_{\bm{n}}$ is the Schur function defined in Eq.~\eqref{eq:Schur_func}.
Then, let us consider, instead of $\diag(\bm{t})^{\otimes n}$ in Eq.~\eqref{eq:inf_coef}, the state $\mathbb{E}_{U_A\sim \text{Haar}(d)}[(U_A\diag(\bm{t})U_A^{\dagger})^{\otimes n}]$, where $\mathbb{E}_{U\sim \text{Haar}(d)}[\cdot]$ denotes the Haar average over the $d$-dimensional unitary group $\mathrm{U}(d)$.  This state commutes with any unitary in the form $U'^{\otimes n}$, $(U'\in \mathrm{U}(d))$, as well as any permutation of the systems, and thus satisfies
\begin{equation}
    P_{\bm{n}}\mathbb{E}_{U_A\sim \text{Haar}(d)}[(U_A\diag(\bm{t}) U_A^{\dagger})^{\otimes n}]P_{\bm{n}}= \alpha_{\bm{n}}(\bm{t})P_{\bm{n}}, \label{eq:direct_sum_rep}
\end{equation}
for coefficients $\alpha_{\bm{n}}(\bm{t})$ satisfying $\sum_{\bm{n}\in Y_n^d} \alpha_{\bm{n}}(\bm{t})=1$.
(This is because $U'^{\otimes n}$ acts as $\pi_{\bm{n}}(U')\otimes I_{{\cal V}_{\bm{n}}}$ on ${\cal U}_{\bm{n}}\otimes {\cal V}_{\bm{n}}$, where $\pi_{\bm{n}}$ is an irreducible representation of $\mathrm{U}(d)$ labeled by $\bm{n}$.)
By taking the trace of both sides, we have
\begin{equation}
    \begin{split}
    &\mathrm{Tr}\!\left[P_{\bm{n}}\mathbb{E}_{U_A\sim \text{Haar}(d)}\Bigl[\bigl(U_A\diag(\bm{t}) U_A^{\dagger}\bigr)^{\otimes n}\Bigr]\right] \\
    &= \alpha_{\bm{n}}(\bm{t}) \dim({\cal U}_{\bm{n}}\otimes {\cal V}_{\bm{n}}).
    \end{split} \label{eq:dim_U_V}
\end{equation}
Since a unitary does not change the spectra, we have
\begin{align}
    &\mathrm{Tr}\!\left(P_{\bm{n}}\mathbb{E}_{U_A\sim \text{Haar}(d)}[(U_A\diag(\bm{t}) U_A^{\dagger})^{\otimes n}]\right) \nonumber \\ 
    &= \mathbb{E}_{U_A\sim \text{Haar}(d)} \left[\mathrm{Tr}(P_{\bm{n}}(U_A\diag(\bm{t}) U_A^{\dagger})^{\otimes n})\right] \\
    &= \mathbb{E}_{U_A\sim \text{Haar}(d)}[\dim {\cal V}_{\bm{n}}\; s_{\bm{n}}(\bm{t})] \\
    &= \dim {\cal V}_{\bm{n}}\; s_{\bm{n}}(\bm{t}). \label{eq:dim_V_schur}
\end{align}
Therefore, from Eqs.~\eqref{eq:dim_U_V} and \eqref{eq:dim_V_schur}, we have
\begin{equation}
    \alpha_{\bm{n}}(\bm{t}) = \frac{s_{\bm{n}}(\bm{t})}{\dim {\cal U}_{\bm{n}}}. \label{eq:alpha_n}
\end{equation}

Now, from Eqs.~\eqref{eq:direct_sum_decomp} and \eqref{eq:def_S}, we have
\begin{equation}
    \begin{split}
    &\mathrm{supp}({\cal S}_{A^n}(P_{\bm{n}} \rho_{A^n} P_{\bm{n}})) \nonumber\\
    &= \begin{cases} \mathrm{supp}(\rho_{{\cal U}_{\bm{n}}}) \otimes \ket{0}\!\bra{0}_{{\cal V}_{\bm{n}}} & \text{on } {\cal U}_{\bm{n}}\otimes {\cal V}_{\bm{n}} \\
        0 & \text{otherwise}
    \end{cases}.
    \end{split}
\end{equation}
Combining the above with Eqs.\eqref{eq:def_S}, \eqref{eq:direct_sum_rep}, and \eqref{eq:alpha_n}, we have
\begin{align}
    &\mathrm{supp}({\cal S}_{A^n}(P_{\bm{n}} \rho_{A^n} P_{\bm{n}}))\nonumber \\ &\leq \frac{{\cal S}_{A^n}(P_{\bm{n}})}{\dim {\cal V}_{\bm{n}}} \\
    \begin{split}
    & =  \frac{1}{\alpha_{\bm{n}}(\bm{t})\dim {\cal V}_{\bm{n}}} \\
    & \quad \times {\cal S}_{A^n}(P_{\bm{n}}\mathbb{E}_{U_A\sim \text{Haar}(d)}[(U_A\diag(\bm{t}) U_A^{\dagger})^{\otimes n}]P_{\bm{n}})
    \end{split}\\
    \begin{split}
    &\leq \frac{\dim{\cal U}_{\bm{n}}}{s_{\bm{n}}(\bm{t})\dim {\cal V}_{\bm{n}}} \\
    & \quad \times {\cal S}_{A^n}(\mathbb{E}_{U_A\sim \text{Haar}(d)}[(U_A\diag(\bm{t}) U_A^{\dagger})^{\otimes n}]). 
    \end{split}\label{eq:permutation_iid_bound}
\end{align}
We thus obtain the following from Eqs.~\eqref{eq:direct_sum_decomp}, \eqref{eq:def_S}, \eqref{eq:def_R}, and \eqref{eq:permutation_iid_bound}.
\begin{lemma}\label{lem:lem_for_theo_1}
    Consider a composite system $A^nE$ associated with a Hilbert space ${\cal H}_A^n\otimes {\cal H}_E$ with $\dim{\cal H}_A=d$. 
    Then, there exists a pair $({\cal S}_{A^n},{\cal R}_{A^n})$ of CPTP maps that satisfies the following conditions (i), (ii), and (iii).
    \medskip

    \noindent (i)~For any density operator $\rho_{A^n E}$ and for any $\bm{n}\in Y_n^d$, we have 
    \begin{equation}
        {\cal S}_{A^n}\otimes \mathrm{Id}_E(P_{\bm{n}}\rho_{A^n E}P_{\bm{n}}) = P_{\bm{n}} {\cal S}_{A^n}\otimes \mathrm{Id}_E(\rho_{A^n E})P_{\bm{n}}.
    \end{equation}
 
    \noindent (ii)~For any $\rho_{A^n E}$ that is invariant under permutation of $n$ subsystems of $A^n$, the pair $({\cal S}_{A^n},{\cal R}_{A^n})$  satisfies
    \begin{align}
        {\cal R}_{A^n}\circ {\cal S}_{A^n}\otimes \mathrm{Id}_E(\rho_{A^n E})=\rho_{A^n E}. 
    \end{align}

    \noindent (iii)~For any $\rho_{A^n}$ that is invariant under permutation of $n$ subsystems of $A^n$ and for any $\bm{n}\in Y_n^d$, we have
        \begin{equation}
            \begin{split}
            &\mathrm{supp}({\cal S}_{A^n}(P_{\bm{n}}\rho_{A^n}P_{\bm{n}})) \\
            &\leq \min_{\bm{t}=(t_1,\ldots,t_d)\geq 0: \sum_{i=1}^d t_i=1} \frac{\dim {\cal U}_{\bm{n}}}{s_{\bm{n}}(\bm{t})\dim {\cal V}_{\bm{n}}}  \\
            &\qquad \times {\cal S}_{A^n}\!\Bigl(\mathbb{E}_{U_A \sim {\rm Haar}(d)}\Bigl[\bigl(U_A\,\diag(\bm{t})\,U_A^{\dagger}\bigr)^{\otimes n}\Bigr]\Bigr)\end{split}\label{eq:t_dependence}
        \end{equation} 
\end{lemma}

Since $\bm{t}=\bm{n}/n$ is a suboptimal choice, Theorem~\ref{theo:refined_de_finetti} directly follows from Lemma~\ref{lem:lem_for_theo_1}.  In fact, this choice is fairly good when $n\gg 1$, which is revealed in the following.
Now we move to the proof of Lemma~\ref{lem:ineq_betwee_funcs}.  
First, it is known that the dimension of ${\cal V}_{\bm{n}}$ can be represented as \cite{Fulton1991,Fulton1996}
\begin{equation}
    \dim {\cal V}_{\bm{n}} = \frac{n!}{\prod_{(i, j)} h_{\bm{n}}(i,j)},
\end{equation}
where $h_{\bm{n}}(i, j)$ denotes the hook length of the Young diagram at $(i,j)$, i.e., $i$-th row and $j$-th column.  On the other hand, the dimension of ${\cal U}_{\bm{n}}$ is given by \cite{Fulton1991,Fulton1996}
\begin{equation}
    \dim {\cal U}_{\bm{n}} = \prod_{(i,j)} \frac{d-i+j}{h_{\bm{n}}(i,j)},
\end{equation}
and therefore,
\begin{equation}
    \frac{\dim {\cal U}_{\bm{n}} }{\dim {\cal V}_{\bm{n}} } = \prod_{(i,j)} \frac{d-i+j}{n!} = \frac{n_1!\ldots n_d!}{n!}\prod_{i=1}^{d} {d-i+n_i \choose d-i},\label{eq:ratio_rep_dim}
\end{equation}
which holds even when $n_j=\cdots =n_d = 0$ for $1\leq j\leq d$. 
Next, we use an alternative expression for the Schur function $s_{\bm{n}}(\bm{t})$ defined in Eq.~\eqref{eq:Schur_func}.  Let $T_{\bm{n}}$ be the set of semistandard Young tableaux of shape $\bm{n}$, i.e., Young diagrams filled with integers from $1$ to $d$ non-decreasingly across each row and increasingly down each column.  Then, we have \cite{Fulton1991,Fulton1996}
\begin{equation}
    s_{\bm{n}}(\bm{t}) = \sum_{T_{\bm{n}}} t_1^{w_1} \cdots t_d^{w_d},
\end{equation}
where $w_i$, $i=1,\ldots,d$, denotes the weight of a given Young tableau, i.e., the occurrences of the number $i$ in the Young tableau.  Note that we define $0^0=1$ here.  The largest element in the above summation is given by filling $i$-th row with $i$, which then implies
\begin{equation}
    s_{\bm{n}}(\bm{t}) \geq t_1^{n_1} \cdots t_d^{n_d} = \Pi_{i=1}^n t_i^{n_i}. \label{eq:prob_multinomial}
\end{equation}
Combining this with Eq.~\eqref{eq:ratio_rep_dim}, we have
\begin{align}
    \frac{\dim {\cal U}_{\bm{n}} }{\dim {\cal V}_{\bm{n}}\; s_{\bm{n}}(\bm{t})} \leq \frac{1}{\mathrm{Mult}_{\bm{t}}(\bm{n})} \prod_{i=0}^{d-1} {n+i \choose i}, 
\end{align}
where $\mathrm{Mult}_{\bm{t}}$ is defined in Eq.~\eqref{eq:multinomial_pdf}.
Now, it is clear that the right-hand side is minimized by the choice $\bm{t}=\bm{n}/n$, which is the reason why Eq.~\eqref{eq:support_upper_bound} in Theorem~\ref{theo:refined_de_finetti} may be a fairly tight bound.  Furthermore, $[\mathrm{Mult}_{\bm{n}/n}(\bm{n})]^{-1}=f_c(n,d;\bm{n})$ is bounded from above as
\begin{align}
    f_c(n,d;\bm{n}) &=
    [\mathrm{Mult}_{\bm{n}/n}(\bm{n})]^{-1} \\ &= \frac{n_1!\cdots n_d!}{n!}\left(\frac{n}{n_1}\right)^{n_1} \cdots \left(\frac{n}{n_d}\right)^{n_d} \\ 
    &\leq  \frac{e^{d}\sqrt{n_1 \cdots n_d}}{\sqrt{2\pi n}} \\
    &\leq \frac{n^{(d-1)/2}}{\sqrt{2\pi (d/e^2)^d}} =f_c(n,d), \label{eq:multinomial_lower_bound}
\end{align}
where we used Stirling's inequality in the first inequality and used the relation between the arithmetic mean and the geometric mean in the second inequality.  This proves Eq.~\eqref{eq:ineq_between_classical_funcs} in Lemma~\ref{lem:ineq_betwee_funcs}.  (Here, we prove Eq.~\eqref{eq:ineq_between_classical_funcs} for $\bm{n}\in Y_n^d$, but this can obviously be generalized to $\bm{n}\in Z_n^d$ from the definition of $\mathrm{Mult}_{\bm{t}}(\bm{n})$ in Eq.~\eqref{eq:multinomial_pdf}.)
We thus have
\begin{align}
    f_q(n,d;\bm{n})&=\frac{\dim {\cal U}_{\bm{n}} }{\dim {\cal V}_{\bm{n}}\; s_{\bm{n}}(\bm{n}/n)}\nonumber\\
    &\leq  \frac{n^{\frac{d-1}{2}}}{\sqrt{2\pi (d/e^2)^{d}}}\prod_{i=0}^{d-1} {n+i \choose i} \label{eq:factorial_bound_first}  \\ 
    &\leq \frac{n^{\frac{d-1}{2}}}{\sqrt{2\pi (d/e^2)^{d}}} \prod_{i=0}^{d-1} \frac{(n+d-1)^i}{i!} \label{eq:factorial_bound_second}\\
    & \leq \frac{(n+d-1)^{\frac{(d^2 - 1)}{2}}}{\sqrt{2\pi (d/e^2)^{d}}G(d+1)} = f_q(n,d) ,\label{eq:coefficient_bound}
\end{align}
where $G(n)$ denotes the Barnes G-function defined in Eq.~\eqref{eq:Barnes_G}.
Notice that we used the following trivial inequalities from Eq.~\eqref{eq:factorial_bound_first} to Eq.~\eqref{eq:factorial_bound_second}
\begin{equation}
    {n+i \choose i} = \overbrace{\frac{(n+i)\cdots(n+1)}{i!}}^{i} \leq (n+d-1)^i/i!.
\end{equation}
(Note: the bound ${n+i \choose i}\leq (n+1)^i$ is conventionally used for this context, but this leads to a looser bound in the case $n \gg d$, which is of our interest.)
This proves Eqs.~\eqref{eq:ineq_between_quantum_funcs} in Lemma~\ref{lem:ineq_betwee_funcs},
which completes the proof of Theorem~\ref{theo:refined_de_finetti} and Lemma~\ref{lem:ineq_betwee_funcs}. \qed

Theorem~\ref{theo:refined_de_finetti} may have lots of applications.  One immediate example is the following.  Assume that a positive operator $\Pi$ satisfies $\mathrm{Tr}(\Pi\rho^{\otimes n}) \leq \mathcal{O}(\exp(-n))$ for any $\rho$.  Such a positive operator $\Pi$ frequently appears in a quantum key distribution or a quantum state verification.  For such a scenario, we have, for any permutation-invariant state $\rho_{A^n}$,
\begin{align}
    \begin{split}
    \mathrm{Tr}[\Pi\rho_{A^n}] &\leq 
    f_q(n,d)\sum_{\bm{n}\in Y_n^d} \mathrm{Tr}[P_{\bm{n}}\rho_{A^n}] \\
    &\quad \times \mathrm{Tr}\!\left(\Pi\,\mathbb{E}_{U\sim \text{Haar}(d)}\Bigl[\Bigl(U\diag\Bigl(\frac{\bm{n}}{n}\Bigr) U^{\dagger}\Bigr)^{\otimes n}\Bigr]\right), \end{split}\\
    &\leq f_q(n,d)\sum_{\bm{n}\in Y_n^d} \mathrm{Tr}[P_{\bm{n}}\rho_{A^n}]\, \sup_{\rho}\mathrm{Tr}(\Pi \rho^{\otimes n}), \\
    & \leq f_q(n,d)\mathcal{O}(\exp(-n)),
\end{align}
and thus almost the same exponential bound holds (up to a polynomial $f_q(n,d)$).
The above inequality will be used later.


\subsection{Generalization to the case with symmetry restrictions}
In this section, we generalize the result of the previous section to the case in which a local $d$-dimensional system has an additional symmetry.  
There are lots of relevant situations in quantum information applications in which a local system has a symmetry.  A classical-quantum state, for example, can be regarded as a quantum state having a $\mathrm{U}(1)$ symmetry in a part of a system.
To make it more concrete, we here derive a tighter upper bound on a permutation-invariant state by i.i.d.~states when a local system has an additional symmetry. 
Let $G$ be a group that acts as a symmetry of local system $A$ whose Hilbert space is ${\cal H}_A$.
Then, without loss of generality, ${\cal H}_A$ can be decomposed into the direct sum given in Eq.~\eqref{eq:decomp_local_system_relabel} on which the representation $\pi$ of the group $G$ acts as in Eq.~\eqref{eq:rep_of_local_symmetry}. 
Any state $\rho_{AE}^G$ that is invariant under the action of the group $G$ on the system $A$ should have the form
\begin{equation}
    \rho_{AE}^G = \bigoplus_{j=1}^k\mathrm{Tr}(Q_j\rho^G) \frac{I_{{\cal H}_j^{R}}}{r_j}\otimes \rho_{{\cal H}_j^{D}E}, \label{eq:symmetric_single}
\end{equation}
with a set of density operators $\{\rho_{{\cal H}_j^{D}E}\}_{j=1}^k$, where $Q_j$ denotes the projection onto the subspace ${\cal H}_j^R\otimes {\cal H}_j^D$.  
Since a $\pi^{\otimes n}(G^{\times n})$-invariant state ${\rho}_{A^nE}^G$ also has this direct-sum structure in each local system $A$ of $A^n$, it has a direct-sum decomposition into operators acting on ${\cal H}_{\bm{j}}^R\otimes {\cal H}_{\bm{j}}^D$ for $\bm{j}\coloneqq(j_1, \ldots, j_n)\in\{1,\ldots,k\}^{\times n}$ given in Eq.~\eqref{eq:decomp_H_bm_j}.
Let $Q_{\bm{j}}$ be a projection operator onto this subspace, i.e.,
\begin{equation}
    Q_{\bm{j}} = Q_{j_1} \otimes \cdots \otimes Q_{j_n}, \label{eq:def_of_Q_bm_j}
\end{equation}
where $Q_{\bm{j}}Q_{\bm{j}'}=\delta_{\bm{j}\bm{j}'}Q_{\bm{j}}$.  Then, the $\pi^{\otimes n}(G^{\times n})$-invariant state ${\rho}_{A^nE}^G$ is decomposed as 
\begin{align}
    {\rho}_{A^nE}^G &= \sum_{\bm{j}\in\{1,\ldots,k\}^{\times n}} Q_{\bm{j}} {\rho}_{A^n E}^G Q_{\bm{j}}\\ 
    &= \bigoplus_{\bm{j}\in\{1,\ldots,k\}^{\times n}} \frac{\mathrm{Tr}[Q_{\bm{j}}\rho_{A^n E}^G]}{\dim {\cal H}_{\bm{j}}^R}I_{{\cal H}_{\bm{j}}^R}\otimes \rho_{{\cal H}_{\bm{j}}^D E} 
    \label{eq:classical_type_decomp}
\end{align} 
with the set of density operators $\{\rho_{{\cal H}_{\bm{j}}^D E}\}$ for $\bm{j}\in\{1,\ldots,k\}^{\times n}$. 

Now we assume that the state $\rho_{A^n E}^G$ is also invariant under the permutation of $n$ systems of $A^n$.  Due to the decomposition of the Hilbert space ${\cal H}_A$ as in Eq.~\eqref{eq:decomp_local_system_relabel}, the unitary representation $V_{\sigma}$ of the permutation $\sigma\in S_n$ acts as $V_{\sigma}=V_{\sigma}^R\otimes V_{\sigma}^D$, where $V_{\sigma}^{R(D)}$ acts on the Hilbert space $\bigl(\bigoplus_{j=1}^k{\cal H}_j^{R(D)}\bigr)^{\otimes n}$.  Note that the Hilbert space $\bigl(\bigoplus_{j=1}^k{\cal H}_j^{R}\bigr)^{\otimes n}\otimes \bigl(\bigoplus_{j=1}^k{\cal H}_j^{D}\bigr)^{\otimes n}$ on which $V_{\sigma}$ acts is larger than ${\cal H}_A^{\otimes n}$, but it contains ${\cal H}_A^{\otimes n}$ as an invariant subspace. 
Now, we define $\bm{\chi}$, ${\cal T}(\bm{m})$, $\bm{j}_0(\bm{m})$, and $\sigma_{\bm{j}}$ as in Eqs.~\eqref{eq:def_bm_chi}, \eqref{eq:def_type_set}, \eqref{eq:def_bm_j_0}, and \eqref{eq:def_sigma_bm_j}, respectively.  Then, we have 
\begin{equation}
    Q_{\bm{j}_0(\bm{m})}V_{\sigma_{\bm{j}}}^{\dagger}V_{\sigma_{\bm{j}'}}Q_{\bm{j}_0(\bm{m})} = \delta_{\bm{j}\bm{j}'}Q_{\bm{j}_0(\bm{m})}, \label{eq:kronecker_delta_Q}
\end{equation}
where $\delta_{\bm{j}\bm{j}'}$ denotes the Kronecker delta.
From $[V_{\sigma_{\bm{j}}}, \rho_{A^nE}^G]=0$ and Eq.~\eqref{eq:classical_type_decomp}, we have 
\begin{equation}
    \rho_{{\cal H}_{\bm{j}}^D E} = V_{\sigma_{\bm{j}}}^D\, \rho_{{\cal H}_{\bm{j}_0(\bm{m})}^D E}\, (V_{\sigma_{\bm{j}}}^D )^{\dagger},
\end{equation}
for any $\bm{j}\in{\cal T}(\bm{m})$.  This leads to the alternative expression of ${\rho}_{A^nE}^G$ as 
\begin{equation}
    \begin{split}
    {\rho}_{A^nE}^G &= \bigoplus_{\bm{m}\in Z_n^k} \frac{\mathrm{Tr}[P_{\bm{m}}\rho_{A^n E}^G]}{r_{\bm{m}}} \\ 
    &\qquad \quad \bigoplus_{\bm{j}\in{\cal T}(\bm{m})} V_{\sigma_{\bm{j}}} \Bigl(I_{{\cal H}_{\bm{j}_0(\bm{m})}^R}\otimes \rho_{{\cal H}_{\bm{j}_0(\bm{m})}^D E}\Bigr) V_{\sigma_{\bm{j}}}^{\dagger},
    \end{split}\label{eq:alternative_decomp}
\end{equation}
where 
\begin{equation}
    P_{\bm{m}}\coloneqq \sum_{\bm{j}\in{\cal T}(\bm{m})} Q_{\bm{j}}
\end{equation}
and 
\begin{equation}
    r_{\bm{m}}\coloneqq |{\cal T}(\bm{m})|\prod_{j=1}^k r_j^{m_j}.  
\end{equation}
Note, for any $\bm{j}\in{\cal T}(\bm{m})$, we have $\dim{\cal H}_{\bm{j}}^R = \prod_{j=1}^k r_j^{m_j}$.

Now, we define a CP map $s[\bm{m}]:{\cal B}({\cal H}_A^{\otimes n})\to {\cal B}({\cal H}_{\bm{j}_0(\bm{m})}^D)$ as follows:
\begin{align}
    s[\bm{m}](\rho) &\coloneqq \!\sum_{\bm{j}\in{\cal T}(\bm{m})}\!(V_{\sigma_{\bm{j}}}^D)^{\dagger}\mathrm{Tr}^{(\bm{j})}_{{\cal H}_{\bm{j}}^R}\bigl(Q_{\bm{j}}\rho Q_{\bm{j}}|_{{\cal H}_{\bm{j}}^{R}\otimes {\cal H}_{\bm{j}}^{D}}\bigr)V_{\sigma_{\bm{j}}}^D. \label{eq:def_small_s} 
\end{align}
Moreover, let $r[\bm{m}]:{\cal B}({\cal H}_{\bm{j}_0(\bm{m})}^D)\to {\cal B}({\cal H}_A^{\otimes n})$ be a CPTP map defined as 
\begin{equation}
    r[\bm{m}](\rho) = \frac{1}{r_{\bm{m}}} \bigoplus_{\bm{j}\in{\cal T}(\bm{m})} I_{{\cal H}_{\bm{j}}^R} \otimes V_{\sigma_{\bm{j}}}^D \rho  (V_{\sigma_{\bm{j}}}^D)^{\dagger}.
\end{equation}
Then, for any state $\rho^G_{A^nE}$ that is invariant under $\pi^{\otimes n}(G^{\times n})$ as well as permutation of $n$ systems of $A^n$, we have 
\begin{equation}
    s[\bm{m}] \otimes \mathrm{Id}_E(\rho^G_{A^nE}) = \mathrm{Tr}[P_{\bm{m}}\rho^G_{A^nE}] \rho_{{\cal H}_{\bm{j}_0(\bm{m})}^D E} \label{eq:projection_to_bm_m}
\end{equation}
and 
\begin{equation}
    \bigoplus_{\bm{m}\in Z_n^k} (r[\bm{m}]\circ s[\bm{m}])\otimes \mathrm{Id}_E (\rho^G_{A^nE}) = \rho^G_{A^nE}. \label{eq:recovery_small}
\end{equation}

For each $\bm{m}=(n_1,\ldots,n_k)\in Z_n^k$, let us pick up a permutation $\sigma\in S_{n_1}\times\cdots\times S_{n_k} \subset S_n$.  Then, we have $[V_{\sigma},\rho_{A^nE}^G]=0$ and $[V_{\sigma},Q_{\bm{j}_0(\bm{m})}]$, which thus leads to 
\begin{equation}
[V_{\sigma},\rho_{{\cal H}_{\bm{j}_0(\bm{m})}^D E}]=0.\label{eq:commutativity_V_sigma}
\end{equation}
Let us recall that $({\cal H}_{j}^D)^{\otimes n_j} = \bigoplus_{\bm{n}_j\in Y_{n_j}^{d_j}}{\cal U}_{\bm{n}_j}\otimes {\cal V}_{\bm{n}_j}$.  Then, we have 
\begin{equation}
    {\cal H}_{\bm{j}_0(\bm{m})}^D = \bigoplus_{\substack{(\bm{n}_1,\ldots,\bm{n}_k) \\ \in Y_{n_1}^{d_1}\times \cdots \times Y_{n_k}^{d_k}}} \bigotimes_{j=1}^k {\cal U}_{\bm{n}_j}\otimes {\cal V}_{\bm{n}_j}. 
\end{equation}
Let $P^D_{\bm{j}_0(\bm{m});\bm{n}_1,\ldots,\bm{n}_k}$ be a projection operator onto the subspace $\bigotimes_{j=1}^k {\cal U}_{\bm{n}_j}\otimes {\cal V}_{\bm{n}_j}$.  Then, from Eq.~\eqref{eq:commutativity_V_sigma}, we have for any $\bm{m}\in Z_n^k$ that
\begin{equation}
    \begin{split}
    &\rho_{{\cal H}_{\bm{j}_0(\bm{m})}^D E} \\
    &= \!\!\bigoplus_{\substack{(\bm{n}_1,\ldots,\bm{n}_k)}}\!\! P^D_{\bm{j}_0(\bm{m});\bm{n}_1,\ldots,\bm{n}_k} \rho_{{\cal H}_{\bm{j}_0(\bm{m})}^D E} P^D_{\bm{j}_0(\bm{m});\bm{n}_1,\ldots,\bm{n}_k}. 
    \end{split}\label{eq:symmetry_decomp_ns}
\end{equation}
Furthermore, from Theorem~\ref{theo:refined_de_finetti} and Eqs.~\eqref{eq:composable_identity}, \eqref{eq:composite_commutativity}, and \eqref{eq:composite_support_bound}, there exists a pair $(\tilde{\cal S}_{{\cal H}_{\bm{j}_0(\bm{m})}^D}, \tilde{\cal R}_{{\cal H}_{\bm{j}_0(\bm{m})}^D})$ of CPTP maps on ${\cal H}_{\bm{j}_0(\bm{m})}^D$ that satisfies the following three conditions (i), (ii), and (iii). 

\noindent (i)~For any $\rho\in{\cal D}({\cal H}_{\bm{j}_0(\bm{m})}^D\otimes {\cal H}_E)$ and for any $\bm{n}_j\in Y_{n_j}^{d_j}$ for $j=1,\ldots,k$, we have 
\begin{equation}
    \begin{split}
    &\tilde{\cal S}_{{\cal H}_{\bm{j}_0(\bm{m})}^D}\otimes \mathrm{Id}_E(P^D_{\bm{j}_0(\bm{m});\bm{n}_1,\ldots,\bm{n}_k}\rho P^D_{\bm{j}_0(\bm{m});\bm{n}_1,\ldots,\bm{n}_k}) \\ &= P^D_{\bm{j}_0(\bm{m});\bm{n}_1,\ldots,\bm{n}_k} \tilde{\cal S}_{{\cal H}_{\bm{j}_0(\bm{m})}^D}\otimes \mathrm{Id}_E(\rho) P^D_{\bm{j}_0(\bm{m});\bm{n}_1,\ldots,\bm{n}_k}. \label{eq:tilde_S_commutativity}
    \end{split}
\end{equation}

\noindent (ii)~For any $\rho\in{\cal D}({\cal H}_{\bm{j}_0(\bm{m})}^D\otimes {\cal H}_E)$ that is $(S_{n_1}\times\cdots\times S_{n_k})$-invariant on ${\cal H}_{\bm{j}_0(\bm{m})}^D$, we have
\begin{equation}
    (\tilde{\cal R}_{{\cal H}_{\bm{j}_0(\bm{m})}^D}\circ \tilde{\cal S}_{{\cal H}_{\bm{j}_0(\bm{m})}^D})\otimes \mathrm{Id}_E(\rho)=\rho. \label{eq:recovery_w_theo_1}
\end{equation}

\noindent (iii)~For any $\rho\in{\cal D}({\cal H}_{\bm{j}_0(\bm{m})}^D)$ that is $(S_{n_1}\times\cdots\times S_{n_k})$-invariant and for any $\bm{n}_j\in Y_{n_j}^{d_j}$ for $j=1,\ldots,k$, we have 
\begin{equation}
    \begin{split}
        &\mathrm{supp}\bigl(\tilde{\cal S}_{{\cal H}_{\bm{j}_0(\bm{m})}^D}(P^D_{\bm{j}_0(\bm{m});\bm{n}_1,\ldots,\bm{n}_k}\rho P^D_{\bm{j}_0(\bm{m});\bm{n}_1,\ldots,\bm{n}_k})\bigr) \\
        &\leq \prod_{j=1}^k f_q(n_j,d_j;\bm{n}_j)\,  \tilde{\cal S}_{{\cal H}_{\bm{j}_0(\bm{m})}^D}\!\biggl(\bigotimes_{j=1}^k \\
        &\qquad \mathbb{E}_{U_j^D\sim \mathrm{Haar}(d_j)}\Bigl[\Bigl(U_j^D \diag\Bigl(\frac{\bm{n}_j}{n_j}\Bigr) (U_j^D)^{\dagger}\Bigr)_{{\cal H}_j^D}^{\otimes n_j} \Bigr]\biggr).
    \end{split} \label{eq:support_bound_j_0}
\end{equation}

Now, define a pair $({\cal S}_{A^n}^G, {\cal R}_{A^n}^G)$ of CPTP maps on ${\cal H}_A^{\otimes n}$ as 
\begin{align}
    {\cal S}_{A^n}^G(\rho) &\coloneqq \bigoplus_{\bm{m}\in Z_n^k}\ket{\varnothing}\!\bra{\varnothing}_{{\cal H}_{\bm{j}_0(\bm{m})}^R} \otimes \bigl(\tilde{\cal S}_{{\cal H}_{\bm{j}_0(\bm{m})}^D}\circ s[\bm{m}](\rho)\bigr), \label{eq:def_S^G}\\
    \begin{split}
    {\cal R}_{A^n}^G(\rho) &\coloneqq \bigoplus_{\bm{m}\in Z_n^k} r[\bm{m}]\circ \tilde{\cal R}_{{\cal H}_{\bm{j}_0(\bm{m})}^D} \circ \mathrm{Tr}^{(\bm{j}_0(\bm{m}))}_{{\cal H}_{\bm{j}_0(\bm{m})}^R} \\
    &\qquad \quad \biggl(\sum_{\bm{j}\in {\cal T}(\bm{m})}V_{\sigma_{\bm{j}}}^{\dagger} Q_{\bm{j}} 
    \rho Q_{\bm{j}} V_{\sigma_{\bm{j}}} \biggl|_{{\cal H}_{\bm{j}_0(\bm{m})}^R\otimes {\cal H}_{\bm{j}_0(\bm{m})}^D}\biggr),
    \end{split} \label{eq:def_R^G}
\end{align}
where $\ket{\varnothing}\!\bra{\varnothing}_{{\cal H}_{\bm{j}_0(\bm{m})}^R}$ is an arbitrary pure state on ${\cal H}_{\bm{j}_0(\bm{m})}^R$.
The map ${\cal S}_{A^n}^G$ is actually trace preserving since $\sum_{\bm{m}\in Z_n^k} s[\bm{m}]$ and $\tilde{\cal S}_{{\cal H}_{\bm{j}_0(\bm{m})}^D}$ are trace preserving.  The map ${\cal R}_{A^n}^G$ is also trace preserving since $r[\bm{m}]$ and $\tilde{\cal R}_{{\cal H}_{\bm{j}_0(\bm{m})}^D}$ are trace preserving and, for any $\rho\in{\cal D}({\cal H}_{A}^{\otimes n})$,
\begin{equation}
    \mathrm{Tr}\left[\sum_{\bm{j}\in {\cal T}(\bm{m})}V_{\sigma_{\bm{j}}}^{\dagger} Q_{\bm{j}} 
    \rho Q_{\bm{j}} V_{\sigma_{\bm{j}}}\right]=\mathrm{Tr}[P_{\bm{m}}\rho].
\end{equation}
Let $\rho_{A^n E}^G$ be $\pi^{\otimes n}(G^{\times n})$-invariant and permutation invariant over $n$ systems of $A^n$.  Then, from Eqs.~\eqref{eq:recovery_small}, \eqref{eq:recovery_w_theo_1}, \eqref{eq:def_S^G}, and \eqref{eq:def_R^G}, we have 
\begin{align}
    &{\cal R}_{A^n}^G\circ {\cal S}_{A^n}^G\otimes \mathrm{Id}_{E}(\rho_{A^n E}^G) \nonumber \\ &= \bigoplus_{\bm{m}\in Z_n^k} \bigl(r[\bm{m}]\circ \tilde{\cal R}_{{\cal H}_{\bm{j}_0(\bm{m})}^D} \circ \tilde{\cal S}_{{\cal H}_{\bm{j}_0(\bm{m})}^D} \circ s[\bm{m}]\bigr)\otimes \mathrm{Id}_{E}(\rho_{A^n E}^G) \\
    &= \rho_{A^n E}^G,
\end{align}
which proves Eq.~\eqref{eq:recovery_symmetric} in Theorem~\ref{theo:de_finetti_with_symmetry}.
Define $P_{\bm{m};\bm{n}_1,\ldots,\bm{n}_k}$ as 
\begin{equation}
    P_{\bm{m};\bm{n}_1,\ldots,\bm{n}_k}=\bigoplus_{\bm{j}\in{\cal T}(\bm{m})} V_{\sigma_{\bm{j}}} \Bigl(I_{{\cal H}_{\bm{j}_0(\bm{m})}^R} \otimes  P^D_{\bm{j}_0(\bm{m});\bm{n}_1,\ldots,\bm{n}_k}\Bigr) V_{\sigma_{\bm{j}}}^{\dagger}.
\end{equation}
Then, combined with Eqs.~\eqref{eq:kronecker_delta_Q}, \eqref{eq:alternative_decomp}, and \eqref{eq:symmetry_decomp_ns}, we have 
\begin{equation}
    \begin{split}
    &\rho_{A^n E}^G \\
    &= \bigoplus_{\bm{m}\in Z_n^k} \bigoplus_{\substack{(\bm{n}_1,\ldots,\bm{n}_k)\\ \in Y_{n_1}^{d_1}\times \cdots\times Y_{n_k}^{d_k}}} P_{\bm{m};\bm{n}_1,\ldots,\bm{n}_k} \rho_{A^n E}^G P_{\bm{m};\bm{n}_1,\ldots,\bm{n}_k},
    \end{split}
\end{equation}
which proves Eq.~\eqref{eq:symmetric_direct_sum_decomp}.  Now, from Eq.~\eqref{eq:def_small_s}, we have 
\begin{equation}
    \begin{split}
    &s[\bm{m}]\otimes \mathrm{Id}_E(P_{\bm{m};\bm{n}_1,\ldots,\bm{n}_k} \rho_{A^n E}^G P_{\bm{m};\bm{n}_1,\ldots,\bm{n}_k}) \\ &= P^D_{\bm{j}_0(\bm{m});\bm{n}_1,\ldots,\bm{n}_k} s[\bm{m}]\otimes \mathrm{Id}_E(\rho_{A^n E}^G) P^D_{\bm{j}_0(\bm{m});\bm{n}_1,\ldots,\bm{n}_k}.
    \end{split} \label{eq:commute_P_j_P_m}
\end{equation}
Then, combining this with Eqs.~\eqref{eq:kronecker_delta_Q}, \eqref{eq:tilde_S_commutativity}, and \eqref{eq:def_S^G}, we prove Eq.~\eqref{eq:symmetric_commutative_proj} in Theorem~\ref{theo:de_finetti_with_symmetry}.
Finally, by combining Eqs.~\eqref{eq:support_bound_j_0} and \eqref{eq:commute_P_j_P_m}, we have 
\begin{equation}
    \begin{split}
        &\mathrm{supp}\Bigl(\tilde{\cal S}_{{\cal H}_{\bm{j}_0(\bm{m})}^D}\bigl(s[\bm{m}](P_{\bm{m};\bm{n}_1,\ldots,\bm{n}_k} \rho_{A^n E}^G P_{\bm{m};\bm{n}_1,\ldots,\bm{n}_k})\bigr)\Bigr) \\
        &\leq \prod_{j=1}^k f_q(n_j,d_j;\bm{n}_j)\, \tilde{\cal S}_{{\cal H}_{\bm{j}_0(\bm{m})}^D}\biggl(\bigotimes_{j=1}^k\\
        &\quad   \mathbb{E}_{U_j^D\sim \mathrm{Haar}(d_j)}\Bigl[\Bigl(U_j^D \diag\Bigl(\frac{\bm{n}_j}{n_j}\Bigr) (U_j^D)^{\dagger}\Bigr)_{{\cal H}_j^D}^{\otimes n_j} \Bigr]\biggr).\label{eq:support_tilde_S_middle}
    \end{split}
\end{equation}
Let $\tau_{\bm{U}}(\bm{n}_1,\ldots,\bm{n}_k)$ be as defined in Eq.~\eqref{eq:def_tau_U}.  Then, from Eq.~\eqref{eq:def_small_s}, we have 
\begin{equation}
    \begin{split}
    &s[\bm{m}]\bigl(\bigl[\tau_{\bm{U}}(\bm{n}_1,\ldots,\bm{n}_k)\bigr]^{\otimes n}\bigr)\\
    &= \mathrm{Mult}_{\bm{m}/n}(\bm{m})\bigotimes_{j=1}^k \Bigl(U_j^D \diag\Bigl(\frac{\bm{n}_j}{n_j}\Bigr) (U_j^D)^{\dagger}\Bigr)_{{\cal H}_j^D}^{\otimes n_j} .
    \end{split}
\end{equation}
Combining this with Eqs.~\eqref{eq:n-dep_coef_classical}, \eqref{eq:def_S^G}, and \eqref{eq:support_tilde_S_middle}, we have 
\begin{align}
    \begin{split}
    &\mathrm{supp}\bigl(S_{A^n}^G(P_{\bm{m};\bm{n}_1,\ldots,\bm{n}_k} \rho_{A^n E}^G P_{\bm{m};\bm{n}_1,\ldots,\bm{n}_k})\bigr)\\ 
    &\leq f_c(n,k;\bm{m}) \prod_{j=1}^k f_q(n_j,d_j;\bm{n}_j) \ket{\varnothing}\!\bra{\varnothing}_{{\cal H}_{\bm{j}_0}^R} \\
    &\quad \otimes \mathbb{E}_{\bm{U}\sim \times_{j=1}^k\mathrm{Haar}(d_j)} \Bigl[\tilde{\cal S}_{{\cal H}_{\bm{j}_0(\bm{m})}^D}\circ s[\bm{m}]\Bigl( \\
    & \hspace{3cm} \tau_{\bm{U}}(\bm{n}_1,\ldots,\bm{n}_k)^{\otimes n}\Bigr) \Bigr] 
    \end{split}\\
    \begin{split}
    &\leq f_c(n,k;\bm{m}) \prod_{j=1}^k f_q(n_j,d_j;\bm{n}_j) \\
    &\quad \ \times  S_{A^n}^G\Bigl(\mathbb{E}_{\bm{U}\sim \times_{j=1}^k\mathrm{Haar}(d_j)}[\tau_{\bm{U}}(\bm{n}_1,\ldots,\bm{n}_k)^{\otimes n}]\Bigr),
    \end{split}
\end{align}
which proves Eq.~\eqref{eq:support_bound_symmetric} in Theorem~\ref{theo:de_finetti_with_symmetry}.\qed

\begin{remark}
    Instead of applying Eq.~\eqref{eq:support_upper_bound} in Theorem~\ref{theo:refined_de_finetti} to Eq.~\eqref{eq:support_bound_j_0}, one can apply Eq.~\eqref{eq:t_dependence} in Lemma~\ref{lem:lem_for_theo_1} to obtain an even tighter bound.  However, the inequality in Theorem~\ref{theo:de_finetti_with_symmetry} may be more convenient for practical use. 
\end{remark}

A particularly interesting case is when $G=\mathrm{U}(1)$.  Since all the irreducible representations of $\mathrm{U}(1)$ is one dimensional, $r_j=1$ for any $j$.  Let us further assume that $d_1=\cdots=d_k=d/k$.  This case can be regarded as a classical-quantum state with $k$ being the dimension of the classical system.  Then, the coefficient would be
\begin{equation}
    f_c(n,k) \prod_{j=1}^k f_q(n_j, d_j) \simeq \mathcal{O}\bigl(n^{\frac{d^2/k-1}{2}}\bigr),
\end{equation}
and thus we have approximately power-of-$k$ improvement compared to the case when we would not consider the symmetry, i.e., when applying Theorem~\ref{theo:refined_de_finetti} naively.

\subsection{Concentration inequality for an independent and identical measurement on an adversarial state} \label{sec:iid_measurement}
Now we leave from concentration inequality for a quantum state and consider a concentration inequality for quantum measurement outcomes.
In this section, we consider the case where independent and identical measurements ${\cal M}^{\otimes n}$ are performed on an adversarial state $\rho$.  Let ${\cal M}^{\otimes n}(\rho):{\cal X}^{\times n}\to[0,1]$ be a probability mass function as defined in Theorem~\ref{theo:reduction_classical_iid}, and let $\mathrm{P}_{\rho}: Z_n^k\to [0,1]$ be another probability mass function defined as
\begin{equation}
    \mathrm{P}_{\rho}(\bm{m})\coloneqq \sum_{\bm{i}\in{\cal T}(\bm{m})}{\cal M}^{\otimes n}(\rho)(\bm{i}).\label{eq:def_P_rho}
\end{equation}
Since we will consider a concentration inequality for the empirical probability of the measurement outcomes, the relevant probability mass function is $\mathrm{P}_{\rho}(\bm{m})$ rather than ${\cal M}^{\otimes n}(\rho)$.  Furthermore, since $\bm{m}$ is invariant under the permutation of measurement outcomes $\bm{i}$, we have
\begin{equation}
    \mathrm{P}_{\rho}(\bm{m}) = \mathrm{P}_{\tilde{\rho}}(\bm{m}),\label{eq:P_rho_equal_P_rho_tilde}
\end{equation}
where $\tilde{\rho}$ is the permutation-symmetrized version of $\rho$.  Thus, we consider $\tilde{\rho}$ instead of $\rho$ in the following analysis.
Due to the permutation invariance, we can apply Theorem~\ref{theo:de_finetti_with_symmetry} to ${\cal M}^{\otimes n}(\tilde{\rho})$ with $r_j=d_j=1$ for $j=1,\ldots,k$ and obtain
\begin{equation}
    {\cal M}^{\otimes n}(\tilde{\rho})(\bm{i})= \sum_{\bm{m}\in Z_n^k} \mathrm{P}_{\tilde{\rho}}(\bm{m}) \sum_{\bm{j}\in{\cal T}(\bm{m})} \frac{\delta_{\bm{j}\bm{i}}}{|{\cal T}(\bm{m})|}, \label{eq:iid_measurement}
\end{equation}
for any $\bm{i}\in{\cal X}^{\times n}$. 
What we want to know is a restriction imposed on the probability $\mathrm{P}_{\tilde{\rho}}(\bm{m})$ from the fact that the independent and identical measurements are performed on the permutation-invariant state.  
To know this, let $\mathcal{P}$ be the probability simplex, i.e., the set of probability vectors $\bm{p}=(p_1,\ldots,p_k)$ with $p_i\geq 0$ and $\sum_{i=1}^k p_i=1$, and $\mathcal{Q}\subseteq \mathcal{P}$ be its subset defined as
\begin{equation}
    \mathcal{Q} = \{\bm{p}\in{\cal P}:\exists\rho\in{\cal D}({\cal H}) \text{ s.t.\ } \forall i\in{\cal X}, p_i={\cal M}(\rho)(i)\}.
\end{equation}
Then, for a particular sequence $\bm{i}=(i_1,\ldots,i_{n})$ of the measurement outcomes, a conditional probability of the $n$-th outcome $i_n$ conditioned on any sequence $\bm{i}'=(i_1,\ldots,i_{n-1})$ of the other $n-1$ outcomes should be in $\mathcal{Q}$ due to the definition of $\mathcal{Q}$.  Let $\bm{l}=(n'_1,\ldots,n'_k)$ with $n'_j=\chi'_j(\bm{i}')$, where $\bm{\chi}':{\cal X}^{\times (n-1)}\to Z_{n-1}^k$ is defined similarly to Eq.~\eqref{eq:def_bm_chi} for $(n-1)$ tuple, and let $\bm{l}[j]\in Z_n^k$ be defined as a $k$-tuple that adds one to $j$-th element of $\bm{l}$, i.e., $\bm{l}[j]=(n'_1,\ldots,n'_{j-1},n'_{j}+1,n'_{j+1},\ldots,n'_k)$.  By applying the above restriction on the conditional probability to Eq.~\eqref{eq:iid_measurement}, we have
\begin{equation}
    \begin{split}
    \left(\sum_{j=1}^k \frac{\mathrm{P}_{\tilde{\rho}}(\bm{l}[j])}{|{\cal T}(\bm{l}[j])|}\right)^{-1}\!\!\left(\frac{\mathrm{P}_{\tilde{\rho}}(\bm{l}[1])}{|{\cal T}(\bm{l}[1])|},\ldots,\frac{\mathrm{P}_{\tilde{\rho}}(\bm{l}[k])}{|{\cal T}(\bm{l}[k])|}\right) 
    \in \mathcal{Q}.
    \end{split}\label{eq:condition_in_Q}
\end{equation}
(Note, if only this condition is satisfied, the following argument holds for any classical probability mass function as well.)
This restriction on $\mathrm{P}_{\tilde{\rho}}(\bm{m})$ for each $\bm{m}$ is strict, but is not very convenient for the later analysis.  So, we consider a looser version for this that is easier to analyze.

Let ${\cal R}\coloneqq \{(p_1,\ldots,p_k)\in {\cal P}:\sum_{j=1}^{k}a_jp_j \leq b\}$ be a convex subset of ${\cal P}$ with an affine boundary such that ${\cal Q} \subseteq {\cal R}$ (see Fig.~\ref{fig:simplex}).  
\emph{We restrict our attention to the case where the complement ${\cal P}\setminus{\cal R}$ of ${\cal R}$ contains only one probability vector that corresponds to having a deterministic outcome}, say $\bm{p}_1\coloneqq(1,0,\ldots,0)$.
This may lead to a looser bound when applied to a specific problem or limit the applicability of the obtained bound, but as can be shown later, it still has a wide range of applications.
The coefficients $\{a_j\}_{j=1}^{k}$ that defines the set ${\cal R}$ in this case should satisfy $a_1>b$, $a_j\leq b$ for $j=2,\ldots,k$.  By redefining $b_j = -(a_j-b)/(a_1-b) (\geq 0)$ for $j=2,\ldots,k$, we have
\begin{align}
    {\cal R}&= \left\{(p_1,\ldots,p_k)\in {\cal P}:p_1 -\biggl(\sum_{j=2}^k b_j p_j\biggr) \leq 0\right\} \\
    &= \left\{\bm{p}=(p_1,\ldots,p_k)\in {\cal P}:\bm{u}\cdot\bm{p}\leq 0\right\},\label{eq:region_R}
\end{align} 
where $\bm{u}\coloneqq (1,-b_2,\ldots,-b_k)$ is as defined in the theorem statement.
In this way, we obtain the expression of ${\cal R}$ in Eq.~\eqref{eq:def_curly_R}.  In other words, Eq.~\eqref{eq:def_curly_R} is defined to satisfy the condition on the complement ${\cal P}\setminus{\cal R}$ stated above.  We also define ${\cal R}'$ as in the theorem statement, i.e.,
${\cal R}'\coloneqq \{\bm{p}\in{\cal P}:\bm{u}\cdot\bm{p} \leq \frac{\sqrt{2}\|\bm{u} - (\bm{v}\cdot\bm{u})\bm{v}\|}{n}\}$, where 
$\bm{v}\coloneqq (1,\ldots,1)/\sqrt{k}$ is as defined in the theorem statement.
Since ${\cal Q}\subseteq{\cal R}$ by assumption, we have from Eq.~\eqref{eq:condition_in_Q} that 
\begin{equation}
    \begin{split}
    \left(\sum_{j=1}^k \frac{\mathrm{P}_{\tilde{\rho}}(\bm{l}[j])}{|{\cal T}(\bm{l}[j])|}\right)^{-1}\!\!\left(\frac{\mathrm{P}_{\tilde{\rho}}(\bm{l}[1])}{|{\cal T}(\bm{l}[1])|},\ldots,\frac{\mathrm{P}_{\tilde{\rho}}(\bm{l}[k])}{|{\cal T}(\bm{l}[k])|}\right) \in \mathcal{R}.
    \end{split}\label{eq:iid_measurement_w_cons}
\end{equation}
Then, with this loosened constraint, we will find an upper bound on $\mathrm{P}_{\tilde{\rho}}(\bm{m})$ for an arbitrary frequency distribution $\bm{m}=(n_1,\ldots,n_k)\in Z_n^k$ of the outcomes. 

As a first step, we will show that given an arbitrary probability vector $\bm{p}\in{\cal R}$, any probability vector $\bm{q}\in{\cal P}$ with $\|\bm{p}-\bm{q}\|\leq \sqrt{2}/n$ is contained in ${\cal R}'$.  From the Cauchy-Schwarz inequality, we have
\begin{align}
    \bm{u}\cdot(\bm{q}-\bm{p}) &= (\bm{u} - (\bm{v}\cdot\bm{u})\bm{v})\cdot(\bm{q}-\bm{p}) \\
    &\leq \|\bm{u} - (\bm{v}\cdot\bm{u})\bm{v}\|\|\bm{q}-\bm{p}\|\\
    &\leq \frac{\sqrt{2}\|\bm{u}- (\bm{v}\cdot\bm{u})\bm{v}\|}{n}, \label{eq:Schwarz}
\end{align}
where we used $\bm{v}\cdot\bm{p} = \sqrt{k} =\bm{v}\cdot\bm{q}$ in the first equality.
(Note that $\bm{u}-(\bm{v}\cdot\bm{u})\bm{v}$ is a vector projection of $\bm{u}$ onto the affine hull of ${\cal P}$ in $\mathbb{R}^k$.)  
Since $\bm{p}\in{\cal R}$ implies $\bm{u}\cdot\bm{p} \leq 0$, we have from Eq.~\eqref{eq:Schwarz} that 
\begin{equation}
    \bm{u}\cdot\bm{q} \leq \bm{u}\cdot\bm{p} + \frac{\sqrt{2}\|\bm{u}- (\bm{v}\cdot\bm{u})\bm{v}\|}{n} \leq \frac{\sqrt{2}\|\bm{u}- (\bm{v}\cdot\bm{u})\bm{v}\|}{n},
\end{equation}
which then implies that $\bm{q}\in{\cal R}'$ from the definition of ${\cal R}'$.

Now, let $\tilde{\cal P}$ be the set of types for $n$ outcomes defined as 
\begin{equation}
    \tilde{\cal P}\coloneqq \{(n_1,\ldots,n_k)/n\in{\cal P}:\forall j\in\{1,\ldots,k\}, n_j\in\mathbb{Z}\},
\end{equation}
and let $\tilde{\cal R}$ be its subset defined as 
\begin{equation}
    \tilde{\cal R} \coloneqq \tilde{\cal P} \cap ({\cal R}'\setminus{\cal R}).
\end{equation}
We prove that every path in $\tilde{\cal P}$ from $\bm{p}_1=(1,0,\ldots,0)$ to $(0,n_2,\ldots,n_k)/n\in{\cal R}$ (for any possible combination $(n_2,\ldots,n_k)$ with $\sum_{j=2}^k n_j=n$) by repeatedly subtracting $1/n$ from the first entry and adding $1/n$ to one of the other entries hits at least one element in $\tilde{\cal R}$.  We define the neighbors of a type $\bm{p}\in\tilde{\cal P}$ to be those types that can be reached by subtracting $1/n$ from one entry of $\bm{p}$ and adding $1/n$ to another entry of $\bm{p}$.
Then, the Euclidean distance between neighboring types is $\sqrt{2}/n$.
If there exists a path from $(1,0,\ldots,0)$ to $(0,n_2,\ldots,n_k)/n$ that does not hit any element of $\tilde{\cal R}$, then that means there exists a pair $(\bm{p},\bm{p}')$ of neighboring probability vectors such that $\bm{p}\in{\cal P}\setminus{\cal R}'$ and $\bm{p}'\in{\cal R}$.  But this contradicts the fact that any $\bm{q}\in{\cal P}$ with $\|\bm{q}-\bm{p}'\|\leq\sqrt{2}/n$ is contained in ${\cal R}'$.  This completes the proof by contradiction.

Now, let us consider the following subnormalized convex mixture $\sigma$ of i.i.d.~probability distributions over ${\cal X}^{\times n}$:
\begin{align}
    \sigma(\bm{i})&=\sum_{\substack{\bm{m}=(n_1,\ldots,n_k)\in Z_n^k\\ :\bm{m}/n\in\tilde{\cal R}}} \mathrm{P}_{\tilde{\rho}}(\bm{m})\prod_{l=1}^n \frac{n_{i_l}}{n} \label{eq:def_of_sigma}\\
    &=\sum_{\substack{\bm{m}=(n_1,\ldots,n_k)\in Z_n^k\\ :\bm{m}/n\in\tilde{\cal R}}} \mathrm{P}_{\tilde{\rho}}(\bm{m})\prod_{j=1}^k \left(\frac{n_{j}}{n}\right)^{\chi_j(\bm{i})},
\end{align}
which is subnormalized since $\sum_{\bm{m}\in Z_n^k:\bm{m}/n\in\tilde{\cal R}}\mathrm{P}_{\tilde{\rho}}(\bm{m})$ may not be unity in general.
Let $\mathrm{Q}_{\sigma}:Z_n^k\to[0,1]$ be a probability mass function associated with $\sigma$ defined as 
\begin{align}
    \mathrm{Q}_{\sigma}(\bm{m})&\coloneqq \sum_{\bm{i}\in{\cal T}(\bm{m})}\sigma(\bm{i})\\
    & = \sum_{\bm{m}'/n\in\tilde{\cal R}} \mathrm{P}_{\tilde{\rho}}(\bm{m}') \mathrm{Mult}_{\bm{m}'/n}(\bm{m}).
\end{align}
From Eqs.~\eqref{eq:coef_classical} and \eqref{eq:multinomial_lower_bound}, we have, for any $\bm{m}\in Z_n^k$ with $\bm{m}/n\in\tilde{\cal R}$,
\begin{equation}
    \mathrm{P}_{\tilde{\rho}}(\bm{m}) \leq f_c(n,k)\mathrm{Q}_{\sigma}(\bm{m}). \label{eq:assumption_type_bound}
\end{equation}
In the following, we inductively prove that $\mathrm{P}_{\tilde{\rho}}(\bm{m})\leq f_c(n,k)\mathrm{Q}_{\sigma}(\bm{m})$ holds for any $\bm{m}\in Z_n^k$ with $\bm{m}/n\in{\cal P}\setminus{\cal R}$.
Let us define $\tilde{\cal R}^{(0)}\coloneqq\tilde{\cal R}$.  Then, for any frequency distribution $\bm{m}$ with its type in $\tilde{\cal R}^{(0)}$, Eq.~\eqref{eq:assumption_type_bound} holds.
Let $\bm{l}\in Z^k_{n-1}$ be an arbitrary frequency distribution such that $\bm{l}[1]/n\in\tilde{\cal P}\setminus\tilde{\cal R}^{(0)}$ and $\bm{l}[j]/n\in\tilde{\cal R}^{(0)}$ for $j=2,\ldots,k$.  (Note that $\bm{l}[1]/n \in {\cal P}\setminus{\cal R}'$ automatically holds if the above conditions are satisfied.)
Such a sequence $\bm{l}$ always exists; if there does not exist such a sequence, then that means there exists $j$ such that $\bm{l}[1]/n\in\tilde{\cal P}\setminus\tilde{\cal R}^{(0)}$ and $\bm{l}[j]/n\notin \tilde{\cal R}^{(0)}$.  If $\bm{l}[j]/n\in\tilde{\cal P}\setminus\tilde{\cal R}^{(0)}$, then we can find a sequence $\bm{l}'$ that satisfies the imposed condition by reducing the first entry of $\bm{l}$ by a positive integer multiple of $1/n$ and add it to the $j$-th entry.  The other possibility is that $\bm{l}[j]/n\in{\cal R}$, but this implies that we can find a path that goes from $\bm{p}_1$ through $\bm{l}[1]/n\in\tilde{\cal P}\setminus\tilde{\cal R}^{(0)} $ and $\bm{l}[j]/n\in {\cal R}$ to $(0,n_2,\ldots,n_k)/n$ (with $\sum_{j=2}^k n_j=n$) without intersecting $\tilde{\cal R}^{(0)}$, which contradicts the fact that every path in $\tilde{\cal P}$ from $\bm{p}_1$ to $(0,n_2,\ldots,n_k)/n$ hits at least one element in $\tilde{\cal R}$.
From the constraint on $\mathrm{P}_{\tilde{\rho}}(\cdot)$ in Eq.~\eqref{eq:iid_measurement_w_cons} and the definition of ${\cal R}$ in Eq.~\eqref{eq:def_curly_R}, we have 
\begin{equation}
    \frac{\mathrm{P}_{\tilde{\rho}}(\bm{l}[1])}{|{\cal T}(\bm{l}[1])|} - \biggl(\sum_{j=2}^{k} b_j \frac{\mathrm{P}_{\tilde{\rho}}(\bm{l}[j])}{|{\cal T}(\bm{l}[j])|}\biggr)\leq 0. \label{eq:type_bound_in_R}
\end{equation}
Furthermore, since $\sigma$ is a mixture of the i.i.d.~probability distributions and is thus invariant under the permutation of $n$ outcomes, it has the same form as the right-hand side of Eq.~\eqref{eq:iid_measurement} with $\mathrm{P}_{\tilde{\rho}}$ replaced with $\mathrm{Q}_{\sigma}$.
Considering again a conditional probability mass function of the $n$-th outcome $i_n$ conditioned on a sequence $\bm{i}'=(i_1, \ldots, i_{n-1})$ of the other $n-1$ outcomes of a particular sequence $\bm{i}=(i_1,\ldots,i_n)$ for $\sigma$, we notice that the resulting conditional probability mass function is an $\bm{i}'$ (or in fact $\bm{l}$) dependent mixture of probability distributions in $\tilde{\cal R}$ since $\sigma$ is a mixture of i.i.d.~probability distributions in $\tilde{\cal R}$.  Combining these facts, we have
\begin{align}
    \begin{split}
        &\left(\sum_{j=1}^k \frac{\mathrm{Q}_{\sigma}(\bm{l}[j])}{|{\cal T}(\bm{l}[j])|}\right)^{-1}\left(\frac{\mathrm{Q}_{\sigma}(\bm{l}[1])}{|{\cal T}(\bm{l}[1])|},\ldots,\frac{\mathrm{Q}_{\sigma}(\bm{l}[k])}{|{\cal T}(\bm{l}[k])|}\right) \\
        &\qquad\in {\cal R}'\setminus{\cal R} \subseteq {\cal P}\setminus{\cal R},
    \end{split}
\end{align}
and thus have from Eq.~\eqref{eq:region_R} that 
\begin{equation}
    \frac{\mathrm{Q}_{\sigma}(\bm{l}[1])}{|{\cal T}(\bm{l}[1])|} - \biggl(\sum_{j=2}^{k} b_j \frac{\mathrm{Q}_{\sigma}(\bm{l}[j])}{|{\cal T}(\bm{l}[j])|}\biggr)\geq 0.\label{eq:iid_bound_out_R}
\end{equation}
 Combining Eqs.~\eqref{eq:assumption_type_bound}, \eqref{eq:type_bound_in_R} and \eqref{eq:iid_bound_out_R}, we have
\begin{equation}
    \mathrm{P}_{\tilde{\rho}}(\bm{l}[1]) \leq f_c(n,k) \mathrm{Q}_{\sigma}(\bm{l}[1]).
\end{equation}
Thus, we have the same inequality as Eq.~\eqref{eq:assumption_type_bound} for $\bm{l}[1]$.  We repeatedly apply the above argument by replacing $\tilde{\cal R}^{(0)}$ with $\tilde{\cal R}^{(1)}\coloneqq \tilde{\cal R}^{(0)}\cup\{\bm{l}[1]/n\}$ and so on, and finally obtains $\mathrm{P}_{\tilde{\rho}}(\bm{m})\leq f_c(n,k)\mathrm{Q}_{\sigma}(\bm{m})$ for any frequency distribution $\bm{m}\in Z_n^k$ with $\bm{m}/n\in{\cal P}\setminus{\cal R}$ by induction.

From Eq.~\eqref{eq:coef_classical} and \eqref{eq:multinomial_lower_bound}, we also have 
\begin{equation}
    \mathrm{P}_{\tilde{\rho}}(\bm{m})\leq \sum_{\bm{m}':\bm{m}'/n\in{\cal R}}f_c(n,k)\,\mathrm{Mult}_{\bm{m}'/n}(\bm{m})
\end{equation}
for any $\bm{m}/n\in{\cal R}$.
Thus, for any $\bm{m}\in Z_n^k$, we have
\begin{align}
    \begin{split}
        \mathrm{P}_{\tilde{\rho}}(\bm{m}) &\leq \sum_{\bm{m}':\bm{m}'/n\in \tilde{\cal R}}f_c(n, k) \mathrm{P}_{\tilde{\rho}}(\bm{m}')\, \mathrm{Mult}_{\bm{m}'/n}(\bm{m}) \\ 
        & \quad +  \sum_{\bm{m}':\bm{m}'/n\in {\cal R}}f_c(n, k) \mathrm{P}_{\tilde{\rho}}(\bm{m}')\, \mathrm{Mult}_{\bm{m}'/n}(\bm{m})
    \end{split} \\
    &= \sum_{\bm{m}':\bm{m}'/n\in{\cal R}'}f_c(n,k) \mathrm{P}_{\tilde{\rho}}(\bm{m}')\, \mathrm{Mult}_{\bm{m}'/n}(\bm{m}).
\end{align}
This inequality immediately implies that for any subset $B$ of $Z_n^k$, we have 
\begin{equation}
    \sum_{\bm{m}\in B}\mathrm{P}_{\tilde{\rho}}(\bm{m}) \leq \max_{\bm{q}\in{\cal R}'} \sum_{\bm{m}\in B}f_c(n,k)\, \mathrm{Mult}_{\bm{q}}(\bm{m}).  \label{eq:upper_mixture_multinomial}
\end{equation}

From this result, we can obtain a concentration inequality for measurement outcomes.
Let $D(\bm{p}\|\bm{q})$ be a Kullback-Leibler divergence given by
\begin{equation}
    \begin{split}
    &D(\bm{p}\|\bm{q})\\
    &\coloneqq \begin{cases}\mathrm{Tr}[\bm{p}\ln\bm{p}-\bm{p}\ln\bm{q}] & \text{supp}(\bm{p}) \subseteq \text{supp}(\bm{q}) \\
        +\infty & \text{Otherwise}
    \end{cases}, 
\end{split}\label{eq:kl_divergence}
\end{equation}
where $\text{supp}(\bm{r})$ denotes the support of the (discrete) probability distribution $\bm{r}$.  Then, the following theorem is known~\cite{Csiszar1984}.
\begin{theorem}[Refined Sanov's theorem for a closed convex set~\cite{Csiszar1984}]
    Let ${\cal A}$ be a non-empty closed convex subset of the $(k-1)$-simplex ${\cal P}$ of probability distributions.  Let $\bm{q}\in{\cal P}$ be a probability distribution, and let $Z_n^k$ be as defined in Eq.~\eqref{eq:def_Z_n^k}.  Then, for a multinomial distribution $\mathrm{Mult}_{\bm{q}}$ of $\bm{q}$, one has 
    \begin{equation}
        \sum_{\bm{m}\in Z_n^k:\bm{m}/n\in{\cal A}}\mathrm{Mult}_{\bm{q}}(\bm{m}) \leq \max_{\bm{p}\in{\cal A}} \exp[-n D(\bm{p}\|\bm{q})].
    \end{equation}
\end{theorem}
\noindent (In the above theorem, if ${\cal A}$ is not restricted to a closed convex subset, then a polynomial factor is multiplied to the right-hand side, which may be more conventional~\cite{Sanov1957,Cover2006}.  However, for our purpose of applying it to quantum measurement scenario, we typically consider a subset with an affine boundary obtained through an operator inequality, and thus the above mentioned restriction on ${\cal A}$ may not lose generality.)
Combining the above theorem with Eq.~\eqref{eq:upper_mixture_multinomial}, we obtain
\begin{align}
    \sum_{\bm{m}/n\in{\cal A}}\mathrm{P}_{\tilde{\rho}}(\bm{m})&\leq \max_{\bm{q}\in{\cal R}'}\sum_{\bm{m}/n\in{\cal A}}f_c(n,k)\, \mathrm{Mult}_{\bm{q}}(\bm{m}) \\
    &\leq \max_{\bm{p}\in{\cal A}}\max_{\bm{q}\in{\cal R}'} f_c(n,k)\exp[-n D(\bm{p}\|\bm{q})].
    \label{eq:bound_using_refined_sanov}
\end{align}
Note that the simultaneous maximization over $\bm{p}\in{\cal A}$ and $\bm{q}\in{\cal R}'$ is a convex optimization problem, and thus the order of maximization does not matter.
Combining this with Eqs.~\eqref{eq:def_P_rho} and \eqref{eq:P_rho_equal_P_rho_tilde}, we prove Theorem~\ref{theo:reduction_classical_iid}.\qed  

Thus, to obtain a tight bound, the regions ${\cal R}'$ and ${\cal A}$ should be chosen so that the KL divergence between the two regions is maximized.

\section{Numerical comparison of the performance with other concentration inequalities}\label{sec:numerical_comparison}
The obtained bound in the previous section has only a polynomial factor compared to the i.i.d.~case, and thus it is asymptotically tight.  However, what one may have an interest in a practical setup is tightness in the finite-size regime.  This is especially important for applications such as QKD.
Here, we compare its tightness with other known concentration inequalities that can deal with non-commutative observables.  One famous example is Azuma's inequality~\cite{Azuma_ineq}, which is conventionally used in the so-called phase error correction approach to prove the security of QKD protocols.  Azuma's inequality does not require permutation invariance unlike Theorem~\ref{theo:refined_de_finetti} and can be applied to arbitrary correlated random variables in combination with the so-called Doob decomposition theorem~\cite{Doob1953}.
Various generalizations of Azuma's inequality have been studied \cite{refined_Azuma_ineq1,refined_Azuma_ineq2}, most of which have tightened the inequality by using the bound on the higher moments of the true distribution.
Recently, Kato has succeeded in tightening Azuma's inequality by using unconfirmed knowledge, i.e., a priori knowledge on the expectation value of random variables \cite{Kato2020}.  The inequality holds even when a priori knowledge is false, and it tightens the inequality to the level of i.i.d.\ case for a binary random variable when a priori knowledge is true, which is extremely useful in applications such as QKD.  Since our method can also be applied to adversarial setups such as QKD, we compare the performance of our new method with these two conventionally used bounds.

In order to avoid the complexity of the analysis, we consider a simple setup.  In the following, we first consider the toy model of estimating the outcomes of a rank-one projective measurement from the outcomes of a slightly tilted projective measurement.  This setup allows us to compare the tightness of the estimation in the finite number of rounds when we apply different concentration inequalities.  Next, we consider a simple quantum key distribution protocol, the three-state protocol, and compare the finite-size key rates obtained through these concentration inequalities.

\subsection{Toy model for statistical estimation} \label{sec:toy_model}
For an $N$-qubit system and positive parameters $\epsilon$ and $r$, we consider two types of binary-outcome measurements performed on a qubit system, type I: $\{\ket{0}\!\bra{0},\ket{1}\!\bra{1}\}$ and type I\!I: $\{\ket{\phi_0}\!\bra{\phi_0}, \ket{\phi_1}\!\bra{\phi_1}\}$, where $\ket{\phi_0}=\sqrt{1-r}\ket{0}+\sqrt{r}\ket{1}$ and $\ket{\phi_1}=\sqrt{r}\ket{0}-\sqrt{1-r}\ket{1}$.  
Both measurements outcome $j\in\{0,1\}$ for the corresponding POVM elements $\ket{j}\!\bra{j}$ and $\ket{\phi_j}\!\bra{\phi_j}$ (i.e., POVM elements for this measurement procedure is given by $\{\frac{1}{2}\ket{0}\!\bra{0},\frac{1}{2}\ket{1}\!\bra{1},\frac{1}{2}\ket{\phi_0}\!\bra{\phi_0},\frac{1}{2}\ket{\phi_1}\!\bra{\phi_1}\}$).  Then, the problem is stated as follows; given a state $\rho$ on $N$-qubit system prepared by an adversary, Alice and Bob perform type I or type I\!I measurement on each of them uniformly randomly.  Alice obtains the results of the type I measurements and Bob obtains those of the type I\!I. From the sum $\hat{B}\coloneqq\sum_{i} \hat{b}_i$ of outcomes $\hat{b}_i\in\{0,1\}$ of the type I\!I measurements, Bob aims at obtaining an upper bound on the sum $\hat{A}\coloneqq \sum_{i} \hat{a}_{i} $ of outcomes $\hat{a}_{i}\in\{0,1\}$ of the type I measurements that Alice has, allowing the failure probability $\epsilon$.

Since the setup of the problem above is invariant under the permutation of $N$-qubit system and uses independent and identical measurements, we can apply Theorems~\ref{theo:refined_de_finetti} and \ref{theo:reduction_classical_iid} as well as Azuma's and Kato's inequalities to this problem.  For a positive parameter $\gamma>0$ and $\Delta>0$, we consider the following region ${\cal A}(\gamma,\Delta)$ as the ``failure region'' and will find an upper bound on the probability to fall in this region:
\begin{equation}
    \begin{split}
    &{\cal A}(\gamma,\Delta) \\
    &\coloneqq \bigl\{(p_1,p_2,p_3): p_1,p_2,p_3\geq0,\, p_1+p_2+p_3=1,\\
    &\hspace{2.65cm} p_1-\gamma p_2 \geq m(\gamma) + \Delta\bigr\},
    \end{split} 
    \label{eq:failure_region}
\end{equation}
where $m(\gamma)$ denotes (an upper bound on) the expectation value and $\Delta$ denotes a deviation from it.
If the probability that a triple $(\hat{A}/N,\hat{B}/N,1-\hat{A}/N-\hat{B}/N)$ of outcomes lies in ${\cal A}(\gamma,\Delta)$ is bounded from above by $\epsilon$, then we can say that the upper bound on $\hat{A}/N$ is given by $\gamma\hat{B}/N + m(\gamma) + \Delta$ allowing the failure probability $\epsilon$.
The term $m(\gamma)$ can be determined as follows.
Let $M_c(\gamma)$ be a self-adjoint operator defined as
\begin{equation}
    M_c(\gamma) = \frac{1}{2}\ket{1}\!\bra{1} - \frac{\gamma}{2} \ket{\phi_1}\!\bra{\phi_1}.
    \label{eq:def_of_M_c}
\end{equation} 
Calculating the largest eigenvalue $m(\gamma)$ of $M_c(\gamma)$, we have the following inequality:
\begin{align}
    M_c(\gamma) &\leq m(\gamma) I, \\
     m(\gamma)&\coloneqq \frac{1-\gamma+\sqrt{(1 - \gamma)^2 + 4r\gamma}}{4}, \label{eq:largest_eigval}
\end{align}
where $I$ denotes the identity operator.  This means that for any density operator $\rho$, we have
\begin{equation}
    \mathbb{E}_{\rho}[\hat{a}_i-\gamma \hat{b}_i] \leq m(\gamma), \label{eq:bound_conditional_expectation}
\end{equation}
which then satisfies Eq.~\eqref{eq:failure_region} in expectation for any $\gamma>0$ and $\Delta>0$.
The reason why $\gamma$ is taken to be positive is now clear.  Since $\ket{1}$ and $\ket{\phi_1}$ has a large overlap, $\hat{A}$ is expected to get larger when $\hat{B}$ gets larger.  Thus, the case $\gamma\leq 0$ is not expected to give a nontrivial bound.

For our purpose, we consider the case in which we fix the failure probability $\epsilon$ and find $\Delta$ as the function of $\gamma$, $N$, and $\epsilon$, which then satisfies $\Delta(\gamma, N, \epsilon)\rightarrow +0$ as $N\rightarrow \infty$, or we fix the deviation $\Delta$ and find $\epsilon$ as the function of $\gamma$, $N$, and $\Delta$, which then satisfies $\epsilon(\gamma, N, \Delta)\rightarrow +0$ as $N\rightarrow \infty$.
That means, we would like to find the function $\Delta(\gamma,N,\epsilon)$ that satisfies
\begin{equation}
    \mathrm{Pr}\left[\left(\frac{\hat{A}}{N},\frac{\hat{B}}{N},1-\frac{\hat{A}}{N}-\frac{\hat{B}}{N}\right)\in{\cal A}\bigl(\gamma,\Delta(\gamma,N,\epsilon)\bigr)\right] \leq \epsilon,
    \label{eq:the_goal}
\end{equation}
or find the function $\epsilon(\gamma,N,\Delta)$ that satisfies
\begin{equation}
    \mathrm{Pr}\left[\left(\frac{\hat{A}}{N},\frac{\hat{B}}{N},1-\frac{\hat{A}}{N}-\frac{\hat{B}}{N}\right)\in{\cal A}\bigl(\gamma,\Delta\bigr)\right] \leq \epsilon(\gamma,N,\Delta). \label{eq:the_goal_2}
\end{equation}

First, we apply Azuma's inequality to the above problem.
For $i\in\{1,\ldots,N\}$, let $\hat{c}_i$ be defined as $\hat{c}_i=\hat{a}_i - \gamma \hat{b}_i$.  For the system $i$ where the type I measurement is chosen, $\hat{b}_i$ is regarded as zero, and for the system $i'$ where the type I\!I measurement is chosen, $\hat{a}_{i'}$ is regarded as zero.  Then, the operator $M_c(\gamma)$ gives the (expectation) value of the random variable $\hat{c}_i$ for a density operator $\rho$ as
\begin{equation}
    \mathbb{E}_{\rho}[\hat{c}_i] = \mathrm{Tr}[M_c(\gamma)\rho].
\end{equation}
Noticing that $\hat{c}_j\in\{0,1,-\gamma\}$, we apply Azuma's inequality \cite{Azuma_ineq} (see also Proposition 2 in Ref.~\cite{Matsuura2023}) to $\hat{c}_j$ to obtain
\begin{equation}
    \begin{split}
    &\mathrm{Pr}\left[\frac{1}{N}\sum_{i}\left(\hat{c}_i - \mathbb{E}[\hat{c}_i \mid \hat{c}_{<i}]\right) \geq t\right] \\
    & \leq \exp\bigl[-2Nt^2/(1+\gamma)^2\bigr], 
    \end{split}\label{eq:application_azuma}
\end{equation} 
where $\hat{c}_{<i}\coloneqq(\hat{c}_0,\ldots,\hat{c}_{i-1})$ with $\hat{c}_0\coloneqq 0$.  Since we have, for any $i\in\{1,\ldots,N\}$,
\begin{equation}
    \mathbb{E}[\hat{c}_i \mid \hat{c}_{<i}] \leq \sup_{\rho}\mathrm{Tr}[M_c(\gamma)\rho] \leq m(\gamma), \label{eq:operator_ineq}
\end{equation}
we obtain, from Eqs.~\eqref{eq:application_azuma} and \eqref{eq:operator_ineq}, the following bound:
\begin{align}
    &\mathrm{Pr}\left[\frac{\hat{A}}{N} - \gamma \frac{\hat{B}}{N} \geq m(\gamma) + (1+\gamma)\sqrt{\frac{\ln(1/\epsilon)}{2N}}\right] \leq \epsilon,
    \label{eq:bound_azuma} \\
    \begin{split}
    &\mathrm{Pr}\left[\frac{\hat{A}}{N} - \gamma \frac{\hat{B}}{N} \geq m(\gamma)  + \Delta\right] \\
    &\leq \exp\bigl[-2N\Delta^2/(1+\gamma)^2\bigr].
    \end{split}
\end{align}
Thus, setting 
\begin{equation}
    \Delta_{\rm azuma}(\gamma,N,\epsilon)=(1+\gamma)\sqrt{\frac{\ln(1/\epsilon)}{2N}}
    \label{eq:func_g_azuma}
\end{equation}
achieves Eq.~\eqref{eq:the_goal}, and setting 
\begin{equation}
    \epsilon_{\rm azuma}(\gamma,N,\Delta)=\exp\bigl[-2N\Delta^2/(1+\gamma)^2\bigr]
    \label{eq:func_eps_azuma}
\end{equation}
 achieves Eq.~\eqref{eq:the_goal_2}.

Next, for the application of Kato's inequality \cite{Kato2020}, we split the random variable $\hat{c}_i$ into $\hat{a}_i$ and $-\gamma\hat{b}_i$, and apply Kato's inequality to $\{\hat{b}_i\}_{i=1}^N$, which will eventually be observed by Bob. (Again, we regard $\hat{b}_i=0$ when the type I measurement is chosen for $i$-th system.)  For $\hat{a}_i$, we simply apply Azuma's inequality (since this cannot be observed by Bob).  Using union bound, we finally obtain the concentration inequalities in the form Eqs.~\eqref{eq:the_goal} and \eqref{eq:the_goal_2}.  In fact, by applying Kato's inequality under the filtration of $\{\hat{a}_{<i},\hat{b}_{<i}\}_{j=1}^N$, we have \cite{Kato2020}
\begin{equation}
    \begin{split}
    &\mathrm{Pr}\left[\sum_{i=1}^N \mathbb{E}[\hat{b}_i\mid \hat{a}_{<i},\hat{b}_{<i}] -  \hat{b}_i \geq \biggl(\beta - \alpha + \frac{2\alpha}{N}\sum_{i=1}^N \hat{b}_i\biggr)\sqrt{N}\right]\\
    &\leq \exp\left(-\frac{2(\beta^2 - \alpha^2)}{(1+\frac{4\alpha}{3\sqrt{N}})^2}\right),
    \end{split} \label{eq:apply_kato}
\end{equation}
where $\alpha, \beta\in\mathbb{R}$ and $\beta\geq 0$.  
We then need to optimize the parameters $\alpha$ and $\beta$ in the inequality with the expected outcome of $\hat{B}$.  In order to obtain a concentration inequality of the type Eq.~\eqref{eq:the_goal}, we optimize $\alpha$ and $\beta$ so that the right-hand side of Eq.~\eqref{eq:apply_kato} is equal to $\epsilon/2$.  Let $B^*$ be the expected outcome of $\hat{B}$.  Then, the good choices of parameters are \cite{Curras2021}
\begin{align}
    \begin{split}
    &\alpha^*(B^*,N,\epsilon/2) \\
    &\coloneqq \frac{27\sqrt{2}N(N - 2B^*)\sqrt{\ln(2/\epsilon)[9B^*(N-B^*)+2N\ln(2/\epsilon)]}}{4(9N+8\ln(2/\epsilon))(9 B^*(N - B^*) + 2N\ln(2/\epsilon))} \\
    &\qquad -\frac{54\sqrt{N}B^*(N-B^*)\ln(2/\epsilon) + 12N^{\frac{3}{2}}(\ln(2/\epsilon))^2}{(9N+8\ln(2/\epsilon))(9 B^*(N - B^*) + 2N\ln(2/\epsilon))}, \end{split} \label{eq:alpha_star}\\
    \begin{split}
    &\beta^*(B^*,N,\epsilon/2)\\
    &\coloneqq\sqrt{[\alpha^*(B^*,N,\epsilon/2)]^2 + \frac{\ln(2/\epsilon)}{2} \left(1 + \frac{4\alpha^*(B^*,N,\epsilon/2)}{3\sqrt{N}} \right)^2}. 
    \end{split}\label{eq:beta_star}
\end{align}
From Eq.~\eqref{eq:apply_kato} and applying Azuma's inequality to $\{\hat{a}_i\}_{i=1}^N$, we have
\begin{equation}
    \begin{split}
    &\mathrm{Pr}\biggl[\sum_{i=1}^N \hat{a}_i - \mathbb{E}[\hat{a}_i\mid \hat{a}_{<i},\hat{b}_{<i}] - \gamma\left(\hat{b}_i - \mathbb{E}[\hat{b}_i\mid \hat{a}_{<i},\hat{b}_{<i}]\right) \\
    &\quad \geq \gamma\sqrt{N} \biggl(\beta^* - \alpha^* + \frac{2\alpha^*}{N}\sum_{i=1}^N \hat{b}_i\biggr) + \sqrt{\frac{N\ln(2/\epsilon)}{2}} \biggr] \\
    & \hspace{7cm} \leq \epsilon,
    \end{split}
\end{equation}
from the union bound, where we abbreviate the dependencies of $\alpha^*$ and $\beta^*$ on $B^*$, $N$, and $\epsilon/2$.
Since 
\begin{equation}
    \mathbb{E}[\hat{a}_i - \gamma\hat{b}_{i}\mid \hat{a}_{<i},\hat{b}_{<i}] \leq \sup_{\rho}\mathrm{Tr}[M_c(\gamma)\rho] \leq m(\gamma)
\end{equation}
again holds, we have
\begin{equation}
    \begin{split}
    &\mathrm{Pr}\Biggl[\frac{\hat{A}}{N} - \gamma \frac{\hat{B}}{N} \geq  m(\gamma)  + \sqrt{\frac{\ln(2/\epsilon)}{2N}}  \\
    & \hspace{2.5cm} + \frac{\gamma}{\sqrt{N}} \biggl(\beta^* - \alpha^* + \frac{2\alpha^* \hat{B}}{N}\biggr)\Biggr]\leq \epsilon.
    \end{split}
    \label{eq:bound_kato}
\end{equation}
Thus, setting
\begin{equation}
    \begin{split}
    &\Delta_{\rm kato}(\gamma,N,\epsilon)  \\
    &= \frac{\gamma}{\sqrt{N}} \biggl[\beta^* - \alpha^*\biggl(1 - \frac{2\hat{B}}{N}\biggr)\biggr]  + \sqrt{\frac{\ln(2/\epsilon)}{2N}}
    \end{split}
    \label{eq:func_g_kato}
\end{equation}
achieves Eq.~\eqref{eq:the_goal} in this case.  

Due to the use of the union bound, the derivation of a concentration inequality of the type Eq.~\eqref{eq:the_goal_2} is not carried out in the same way as the case of Azuma's inequality.  With Corollary~1 in Ref.~\cite{Kato2020} applied to $\{\hat{b}\}_{i=1}^N$ and $\delta$ and $\epsilon$ in Corollary~1 set to $1-2B^*/N$ and $\Delta'\sqrt{N}$, respectively, we have
\begin{equation}
    \begin{split}
    &\mathrm{Pr}\biggl[\frac{1}{N}\sum_{i=1}^N (\mathbb{E}[\hat{b}_i\mid \hat{a}_{<i},\hat{b}_{<i}] - \hat{b}_i) \geq \xi(\hat{B},B^*,\Delta')\Delta'\biggr] \\
    &\qquad \leq \exp\left[-\frac{2N\Delta'^2}{1-(1-2\frac{B^*}{N}-\frac{4}{3}\Delta')^2}\right],
    \end{split} \label{eq:kato_large_deviation}
\end{equation}
where $\xi(\hat{B},B^*,\Delta')$ is defined as
\begin{equation}
    \xi(\hat{B},B^*,\Delta')\coloneqq \frac{1-(1-2\frac{\hat{B}}{N})(1-\frac{B^*}{N}-\frac{4}{3}\Delta')}{1-(1-2\frac{B^*}{N})(1-\frac{B^*}{N}-\frac{4}{3}\Delta')}.
\end{equation}
When $\hat{B}=B^*$, it is unity, i.e., $\xi(B^*,B^*,\Delta')=1$.  Now, given $\Delta$, choose $\Delta'$ so that $\gamma \xi(\hat{B},B^*,\Delta') \Delta' \leq \Delta$.  Then, with Azuma's inequality applied to $\{\hat{a}_i\}_{i=1}^N$, we have
\begin{equation}
    \begin{split}
    &\mathrm{Pr}\biggl[\frac{1}{N}\sum_{i=1}^N (\hat{a}_i - \mathbb{E}[\hat{a}_i\mid \hat{a}_{<i},\hat{b}_{<i}]) \geq \Delta-\gamma\xi(\hat{B},B^*,\Delta')\Delta'\biggr]  \\
    & \qquad \leq \exp\left[-2N(\Delta-\gamma\xi(\hat{B},B^*,\Delta')\Delta')^2\right].
    \end{split} \label{eq:azuma_large_deviation}
\end{equation}
Combining Eqs.~\eqref{eq:bound_conditional_expectation}, \eqref{eq:kato_large_deviation}, and \eqref{eq:azuma_large_deviation}, we have 
\begin{equation}
    \mathrm{Pr}\biggl[\frac{\hat{A}}{N}-\gamma\frac{\hat{B}}{N} \geq m(\gamma) + \Delta \biggr] \leq  \epsilon_{\rm kato}(\gamma,N,\Delta),
\end{equation}
where $\epsilon_{\rm kato}(\gamma,N,\Delta)$ is given by 
\begin{equation}
    \begin{split}
    &\epsilon_{\rm kato}(\gamma,N,\Delta) \\
    &= \min_{0\leq\gamma\xi(\hat{B},B^*,\Delta')\Delta'\leq \Delta} \exp\left[-\frac{2N\Delta'^2}{1-(1-2\frac{B^*}{N}-\frac{4}{3}\Delta')^2}\right]\\
    &\hspace{2.2cm} + \exp\left[-2N(\Delta-\gamma\xi(\hat{B},B^*,\Delta')\Delta')^2\right].
    \end{split} \label{eq:epsilon_kato}
\end{equation}

Third, we consider applying Theorem~\ref{theo:refined_de_finetti} to this problem.  For that, we consider the case of i.i.d.~quantum state first.  
Let ${\cal Q}$ be the convex set of probability mass functions that are realizable by this measurement setup.  Since we estimate the frequency of the event $(\text{I},1)$ from that of $(\text{I\!I},1)$, we need not to distinguish $(\text{I},0)$ and $(\text{I\!I},0)$, and thus we group these outcomes as a single outcome $(\cdot,0)$.
Thus, the set ${\cal Q}$ can be explicitly given as
\begin{equation}
    \begin{split}
    {\cal Q}&\coloneqq \biggl\{\bigl(p(\text{I},1), p(\text{I\!I},1), p(\cdot,0)\bigr): \forall\rho,\; p(\text{I},1)=\frac{\braket{1|\rho|1}}{2}, \\
    &\qquad p(\text{I\!I},1)=\frac{\braket{\phi_1|\rho|\phi_1}}{2}, p(\cdot, 0)=\frac{\braket{0|\rho|0} + \braket{\phi_0|\rho|\phi_0}}{2}\biggr\}.
    \end{split}
    \label{eq:def_of_Q}
\end{equation}
Now, assume that the i.i.d.~state $\rho(\bm{q})^{\otimes n}$ is prepared with $\rho(\bm{q})$ giving the probability mass function $\bm{q}\in{\cal Q}$.  Then, the probability that the sums of the outcomes of these measurements $\hat{A}$ and $\hat{B}$ lies in the region ${\cal A}(\gamma,\Delta_{\rm iid})$ is given from the refined version of the Sanov's theorem \cite{Sanov1957, Csiszar1984} as
\begin{equation}
    \begin{split}
    &\mathrm{Pr}\left[\left(\frac{\hat{A}}{N},\frac{\hat{B}}{N},1-\frac{\hat{A}}{N}-\frac{\hat{B}}{N}\right)\in{\cal A}(\gamma,\Delta_{\rm iid})\middle| \rho(\bm{q})^{\otimes n}\right] \\
    &\leq \max_{\bm{p}\in {\cal A}(\gamma,\Delta_{\rm iid})}\exp[-N D(\bm{p}\|\bm{q})].
    \end{split}
\end{equation}
Then, for any i.i.d.~state $\rho^{\otimes N}$, we have the following bound:
\begin{equation}
    \begin{split}
    &\mathrm{Pr}\left[\left(\frac{\hat{A}}{N},\frac{\hat{B}}{N},1-\frac{\hat{A}}{N}-\frac{\hat{B}}{N}\right)\in {\cal A}(\gamma,\Delta_{\rm iid})\middle|\rho^{\otimes N}\right] \\
    &\leq \max_{\bm{p}\in {\cal A}(\gamma,\Delta_{\rm iid})}\max_{\bm{q}\in{\cal Q}}\exp[-N D(\bm{p}\|\bm{q})].
    \end{split}
    \label{eq:concentration_iid}
\end{equation}
Now, we consider a permutation-invariant state $\rho_{\rm sym}$ and apply Theorem~\ref{theo:refined_de_finetti} to have
\begin{equation}
    \begin{split}
    &\mathrm{Pr}\left[\left(\frac{\hat{A}}{N},\frac{\hat{B}}{N},1-\frac{\hat{A}}{N}-\frac{\hat{B}}{N}\right)\in{\cal A}(\gamma,\Delta_q)\middle| \rho_{\rm sym}\right] \\
    &\leq \max_{\bm{p}\in{\cal A}(\gamma,\Delta_q)}\max_{\bm{q}\in{\cal Q}}f_q(N,2)\exp[-N D(\bm{p}\|\bm{q})].
    \end{split}
    \label{eq:quantum_perm_inv_bound}
\end{equation}
However, since the sum $\hat{A}$ of measurement outcomes is permutation invariant and thus the permutation symmetrization does not change the probability, we can apply the same bound for any state that is not necessarily permutation invariant.  Finally, by choosing the parameter $\Delta_q$ so that the right-hand side of Eq.~\eqref{eq:quantum_perm_inv_bound} is equal to $\epsilon$, we obtain Eq.~\eqref{eq:the_goal}.  (The parameter $\Delta_q$ is thus the function of $\gamma$, $N$, and $\epsilon$.)  On the other hand, if we fix $\Delta_q=\Delta$ and set
\begin{equation}
    \epsilon_q(\gamma,N,\Delta) = \max_{\bm{p}\in{\cal A}(\gamma,\Delta)}\max_{\bm{q}\in{\cal Q}}f_q(N,2)\exp[-N D(\bm{p}\|\bm{q})],
    \label{eq:epsilon_q}
\end{equation}
then we obtain a concentration inequality of the type Eq.~\eqref{eq:the_goal_2}.
For later use, we define $\Delta_{\rm iid}(\gamma,N,\epsilon)$ and $\epsilon_{\rm iid}(\gamma,N,\Delta)$ in the same way using Eq.~\eqref{eq:concentration_iid}, i.e.,
\begin{equation}
    \epsilon_{\rm iid}(\gamma,N,\Delta) = \max_{\bm{p}\in{\cal A}(\gamma,\Delta)}\max_{\bm{q}\in{\cal Q}}\exp[-N D(\bm{p}\|\bm{q})].
    \label{eq:epsilon_iid}
\end{equation}

We finally consider applying Theorem~\ref{theo:reduction_classical_iid} to this problem.  As previously mentioned, an independent and identical measurement is performed on an unknown quantum state and the set of measurement outcomes is given by $\{(\text{I},0), (\text{I},1), (\text{I\!I},0), (\text{I\!I},1)\}$.  Since we do not distinguish $(\text{I\!I},0)$ and $(\text{I\!I},1)$ as mentioned previously, we regard this measurement as three-outcome measurement and set $k=3$.
For the set ${\cal Q}$ of probability mass functions that can be realized by this measurement given in Eq.~\eqref{eq:def_of_Q}, we consider the following set ${\cal R}(\gamma')\supseteq{\cal Q}$:
\begin{equation}
    \begin{split}
    {\cal R}(\gamma') &= \Bigl\{(p(\text{I},1),p(\text{I\!I},1),p(\cdot,0)): \\
    &\qquad \qquad \qquad p(\text{I},1)-\gamma' p(\text{I\!I},1)\leq m(\gamma') \Bigr\}.
    \end{split}
\end{equation}
The fact that this includes ${\cal Q}$ as a subset follows from Eqs.~\eqref{eq:def_of_M_c} and \eqref{eq:largest_eigval}.  
Due to the requirement of Theorem~\ref{theo:reduction_classical_iid} that the complement ${\cal P}\setminus{\cal R}(\gamma')$ of ${\cal R}(\gamma')$ contains only one probability mass function that corresponds to having a deterministic outcome, the parameter $\gamma'$ is chosen to satisfy $0\leq m(\gamma') < 1$.
Likewise, the region ${\cal R}'(\gamma')$ can be defined through Eq.~\eqref{eq:region_sifted_by_N_inverse} by 
\begin{equation}
    \begin{split}
    {\cal R}'(\gamma') &= \biggl\{(p(\text{I},1),p(\text{I\!I},1),p(\cdot,0)): \\
    &\ p(\text{I},1)- \gamma'p(\text{I\!I},1)  \leq m(\gamma') + \frac{\sqrt{2}}{N}\sqrt{\frac{2\gamma'^2+2\gamma'+2}{3}} \biggr\}.
    \end{split}
\end{equation}
Thus, from Eq.~\eqref{eq:bound_using_refined_sanov}, we have
\begin{equation}
    \begin{split}
    &\mathrm{Pr}\left[\left(\frac{\hat{A}}{N},\frac{\hat{B}}{N},1-\frac{\hat{A}}{N}-\frac{\hat{B}}{N}\right)\in{\cal A}(\gamma,\Delta_c)\right] \\
    &\leq \max_{\bm{p}\in{\cal A}(\gamma,\Delta_c)}\min_{\gamma':0\leq m(\gamma')< 1}\max_{\bm{q}\in{\cal R}'(\gamma')}f_c(N,3)\exp[-N D(\bm{p}\|\bm{q})],
    \end{split}
    \label{eq:bound_by_classical_iid}
\end{equation}
where minimization over $\gamma'$ is inserted because it is a free parameter.
Then, by setting the parameter $\Delta_c$ so that the right-hand side of Eq.~\eqref{eq:bound_by_classical_iid} is equal to $\epsilon$, we achieve Eq.~\eqref{eq:the_goal}.  (The parameter $\Delta_c$ is thus the function of $\gamma$, $N$, and $\epsilon$.) 
On the other hand, by setting $\Delta_c=\Delta$ and setting
\begin{equation}
    \begin{split}
    &\epsilon_c(\gamma,N,\Delta)\\
    &=\max_{\bm{p}\in{\cal A}(\gamma,\Delta)}\min_{\gamma':0\leq m(\gamma')< 1}\max_{\bm{q}\in{\cal R}'(\gamma')}f_c(N,3)\exp[-N D(\bm{p}\|\bm{q})],
    \end{split}
    \label{eq:epsilon_c}
\end{equation}
we obtain a concentration inequality of the type Eq.~\eqref{eq:the_goal_2}.

When applied to this specific problem, we can contrast the bounds Eqs.~\eqref{eq:epsilon_q} and \eqref{eq:epsilon_c}.  In Eq.~\eqref{eq:epsilon_q}, the polynomial factor depends on the dimension of the local quantum system that is eventually measured, but the region with which the maximization of the Kullback-Leibler divergence is taken is the convex set ${\cal Q}$ of all the possible probability mass functions realized by the measurement $\{\frac{1}{2}\ket{1}\!\bra{1},\frac{1}{2}\ket{\phi_1}\!\bra{\phi_1},\frac{1}{2}(\ket{0}\!\bra{0}+\ket{\phi_0}\!\bra{\phi_0})\}$ on all the possible local quantum states.  In Eq.~\eqref{eq:epsilon_c} on the other hand, the polynomial factor depends on the number of outcomes of the measurement and is independent of the dimension of the underlying quantum system.  The region with which the maximization of the Kullback-Leibler divergence is taken is, however, larger than ${\cal Q}$.  This may be a price to pay.  When the dimension of the quantum system is small but the number of outcomes of the measurement is large, Eq.~\eqref{eq:epsilon_q} may be smaller.  On the contrary, if the dimension of the quantum system is large (even infinite) but the number of outcomes of the measurement is small, Eq.~\eqref{eq:epsilon_c} may be smaller.

\begin{figure}[t]
    \centering
    \includegraphics[width=0.99\linewidth]{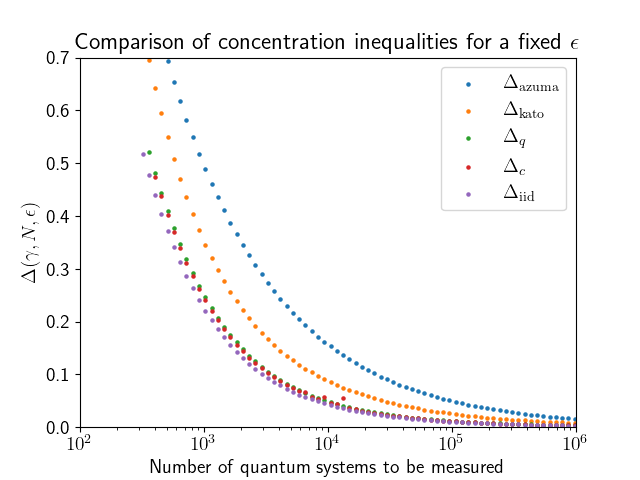}
    \caption{Comparison of the deviation $\Delta(\gamma, N, \epsilon)$ from (an upper bound on) the expectation value in Eqs.~\eqref{eq:failure_region} and \eqref{eq:the_goal} when allowing the failure probability $\epsilon$.  The parameters are chosen to be $\epsilon=10^{-30}$, $r=0.01$, and $B^*=0.01N$, and $\gamma$ is optimized to be $1.662$ so that an unconfirmed upper bound $\gamma B^*/N + m(\gamma)$ on $\hat{A}/N$ is minimized to be $0.02873$.  We further assume that the random variable $\hat{B}$ is equal to the expected value $B^*$.  The functions $\Delta_{\rm azuma}$ and $\Delta_{\rm kato}$ are defined respectively in Eqs.~\eqref{eq:func_g_azuma} and \eqref{eq:func_g_kato}, and the functions $\Delta_q$, and $\Delta_c$ are defined implicitly through Eqs.~\eqref{eq:quantum_perm_inv_bound} and \eqref{eq:bound_by_classical_iid}.  For comparison, we also plot the deviation $\Delta_{\rm iid}$ when the i.i.d.~quantum state is assumed to be given, where $\Delta_{\rm iid}$ is defined implicitly through Eq.~\eqref{eq:concentration_iid}.
    }
    \label{fig:bound_on_hat_A_over_N}    
\end{figure}

We numerically demonstrate the tightness of these bounds in some cases.  The first example is the case that is relevant to the finite-size analysis of the QKD.  In QKD, the parameter $\epsilon$ is fixed and its value is typically chosen to be $\sim 10^{-30}$.  In the QKD application, we have the expected value $B^*$ of $\hat{B}$ from the channel that we use for the transmission, which is the reason why Kato's inequality~\cite{Kato2020} typically gives a tighter bound in the finite-size regime.  For a model, we set $\epsilon=10^{-30}$, $r=0.01$, and $B^*=0.01N$.  The parameter $\gamma$ in ${\cal A}(\gamma,\Delta)$ is chosen so that the expected (unconfirmed) upper bound $\gamma B^*/N+m(\gamma)$ on $\hat{A}/N$ in the limit $N\rightarrow \infty$ is minimized.  In the choice of parameters above, we have $\gamma=1.662$ to give the expected asymptotic upper bound $0.02873$ on $\hat{A}/N$.  We further assume the case in which $\hat{B}$ is exactly equal to $B^*$.  
Under these assumptions, we derive the deviation from the above asymptotic upper bound for finite $N$ allowing the failure probability $\epsilon$ with the four methods described above.
Figure~\ref{fig:bound_on_hat_A_over_N} shows the deviation $\Delta(\gamma, N, \epsilon)$ when varying the number $N$ of quantum systems to be measured.  As the figure suggests, $\Delta_c$ defined through Eq.~\ref{eq:bound_by_classical_iid} is the smallest among four bounds in an adversarial setup (i.e., excluding $\Delta_{\rm iid}$) and with the above particular choices of parameters, except for some points that may be caused by the insufficient optimization.  We can also see that $\Delta_q$ defined through Eq.~\eqref{eq:quantum_perm_inv_bound} gives the comparable bound, and these two ($\Delta_c$ and $\Delta_q$) are as small as that of the i.i.d.~case given by $\Delta_{\rm iid}$.  They are smaller than $\Delta_{\rm kato}$ defined in Eq.~\eqref{eq:func_g_kato}, while $\Delta_{\rm kato}$ is still smaller than the $\Delta_{\rm azuma}$ defined in Eq.~\eqref{eq:func_g_azuma}.

\begin{figure}[t]
    \centering
    \includegraphics[width=0.99\linewidth]{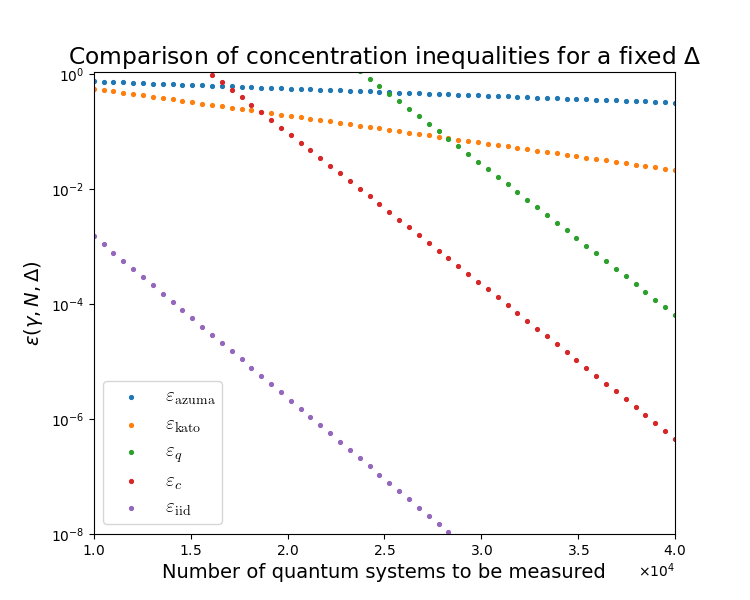}
    \caption{Comparison of the failure probability $\epsilon(\gamma, N, \Delta)$ in Eq.~\eqref{eq:the_goal_2} against the number $N$ of measured quantum systems for a fixed value of the deviation $\Delta$.  The parameters are chosen to be $\Delta=0.01$, $r=0.01$, and $B^*=0.01N$, and $\gamma$ is optimized to be $1.662$ so that an unconfirmed upper bound $\gamma B^*/N + m(\gamma)$ on $\hat{A}/N$ is minimized to be $0.02873$.  We further assume that the random variable $\hat{B}$ is equal to the expected value $B^*$.  The functions $\epsilon_{\rm azuma}$, $\epsilon_q$, $\epsilon_c$, and $\epsilon_{\rm iid}$ are defined in Eqs.~\eqref{eq:func_eps_azuma}, \eqref{eq:epsilon_q}, \eqref{eq:epsilon_c}, and \eqref{eq:epsilon_iid}, respectively.  
    }
    \label{fig:epsilon_vs_N}    
\end{figure}

The next example is the case when we fix the deviation $\Delta$ from the expectation value (that is, an upper bound that holds in the asymptotic limit as shown in Eq.~\eqref{eq:operator_ineq}) to a small value, and see how quickly the failure rate $\epsilon(\gamma, N, \Delta)$ in Eq.~\eqref{eq:the_goal_2} goes to zero. 
We set the deviation $\Delta=0.01$, $r=0.01$ and $B^*=0.01N$, and choose $\gamma=1.662$ to give the asymptotically optimal upper bound $0.02873$ on $\hat{A}/N$ as was done in the previous figure.  We also consider the case in which the observed value $\hat{B}$ is exactly equal to $B^*$.  With these conditions, we plot $\epsilon_{\rm azuma}$ in Eq.~\eqref{eq:func_eps_azuma}, $\epsilon_{\rm kato}$ in Eq.~\eqref{eq:epsilon_kato}, $\epsilon_q$ in Eq.~\eqref{eq:epsilon_q}, $\epsilon_c$ in Eq.~\eqref{eq:epsilon_c}, and $\epsilon_{\rm iid}$ in Eq.~\eqref{eq:epsilon_iid} in Fig.~\ref{fig:epsilon_vs_N}.  
As the figure suggests, the failure probability exponentially decays, but the exponents are very different between Azuma's inequality, Kato's inequality, and ours.
In fact, the convergence speed of our bounds is much quicker than that of Azuma's or Kato's inequality and as quick as that of i.i.d..  Furthermore, $\epsilon_q$, $\epsilon_c$, and $\epsilon_{\rm iid}$ appear to have almost constant gaps in the logarithmic plot.  This gap exactly comes from the difference of the factors $f_q(N,d=2)\sim N^{\frac{3}{2}}$, $f_c(N,k=3)\sim N$, and $1$.  We should note that there are regions in which Azuma's and Kato's inequality gives a better bound than ours when $N$ is small and the failure probability is large.  When we alter the parameters $r$, $M$, and $\Delta$ in our model, then this region varies.  However, our bound overall gives a faster exponential decay.

\subsection{Application to quantum key distribution} \label{sec:three_state}
For the next application, we consider a QKD protocol with qubits.  Even though Bennett-Brassard 1984 protocol~\cite{Bennett1984} may be the most common QKD protocol, its finite-size security analysis can be reduced to a classical random sampling problem~\cite{Tomamichel2012,Hayashi2012,Lim2014} in which one can apply a better concentration inequality than Azuma's and our inequalities.  Therefore, we alternatively consider Phoenix-Barnett-Chefles 2000 (PBC00) protocol \cite{Phoenix2000}, which uses three quantum states forming an equilateral triangle on the $X$-$Z$ plane in the Bloch sphere.  The security of this protocol is proved in the finite-size case against general attacks in Ref.~\cite{Boileau2005} using Azuma's inequality.  In order to make the paper as self-contained as possible, we list the protocol below.
Let $\{\ket{0},\ket{1}\}$ be the $Z$ basis of a qubit.  Let $\{\ket{+},\ket{-}\}$ with $\ket{\pm}\coloneqq(\ket{0}\pm\ket{1})/\sqrt{2}$ be the $X$ basis.  The three states used in PBC00 protocol are $\ket{\psi_1}\coloneqq \cos(2\pi/3)\ket{+}+\sin(2\pi/3)\ket{-}$, $\ket{\psi_2}\coloneqq \cos(4\pi/3)\ket{+}+\sin(4\pi/3)\ket{-}$, and $\ket{\psi_3}\coloneqq \ket{+}$.  Let us also define $\ket{\bar{\psi}_i}$ for $i=0,1,2$ that satisfies $\braket{\psi_i|\bar{\psi}_i}=0$.
Then, the PBC00 protocol is described as follows.

\medskip
\noindent --- {\bf PBC00} ---
\begin{enumerate}
    \item For each of $N$ rounds, Alice uniformly randomly generates a trit $\hat{r}_i$ and a bit $\hat{a}_i$, where $i\in\{1,\ldots,N\}$ denotes the label of the round.  She chooses the pair $\{\ket{\psi_1},\ket{\psi_2}\}$ if $\hat{r}_i=0$, $\{\ket{\psi_2},\ket{\psi_3}\}$ if $\hat{r}_i=1$, and $\{\ket{\psi_3},\ket{\psi_1}\}$ if $\hat{r}_i=2$.  If $\hat{a}_i=0(1)$, she sends the first (second) state of the chosen pair to Bob.
    \item For each qubit received in the $i$-th round, Bob performs a measurement described by the POVM $\{\frac{2}{3}\ket{\bar{\psi}_1}\!\bra{\bar{\psi}_1}, \frac{2}{3}\ket{\bar{\psi}_2}\!\bra{\bar{\psi}_2}, \frac{2}{3}\ket{\bar{\psi}_3}\!\bra{\bar{\psi}_3}\}$.  
    \item After the $N$ rounds of communication, Alice announces the sequence of trits $\{\hat{r}_i\}_{i=1}^N$.  Bob defines $\hat{b}_i=0(1)$ if $\hat{r}_i=0$ and his measurement outcome in the $i$-th round is $\ket{\bar{\psi}_2}$ ($\ket{\bar{\psi}_1}$), if $\hat{r}_i=1$ and his measurement outcome in the $i$-th round is $\ket{\bar{\psi}_3}$ ($\ket{\bar{\psi}_2}$), and if $\hat{r}_i=2$ and his measurement outcome in the $i$-th round is $\ket{\bar{\psi}_1}$ ($\ket{\bar{\psi}_3}$).  All other events are regarded as inconclusive.  Bob announces whether his measurement is inconclusive or not for each $i$.  Alice and Bob keep $\hat{a}_i$ and $\hat{b}_i$, respectively, when Bob's outcome in the $i$-th round is conclusive.  Let $\hat{N}_{\rm con}$ be the number of conclusive rounds.
    \item For each element of the $\hat{N}_{\rm con}$-bit sequence that Alice keeps in the previous step, Alice uniformly randomly chooses the label ``test'' or ``sift'' and announces it.  Let $\hat{N}_{\rm test}$ and $\hat{N}_{\rm sift}$ be the number of test and sift bits, respectively, which satisfies $\hat{N}_{\rm con}=\hat{N}_{\rm test} + \hat{N}_{\rm sift}$.  Alice and Bob announce all the bits labeled by ``test'' and estimate the bit error rate.  From the estimated bit error rate, they perform a bit error correction on the sifted key, i.e., $\hat{N}_{\rm sift}$-bit sequence labeled by ``sift''.  Let $\hat{N}_{\rm KC}$ be the number of pre-shared secret keys consumed during the bit error correction.
    \item Alice and Bob agree on the amount $\hat{N}_{\rm PA}$ of privacy amplification and perform the linear hash function to obtain the final key.
\end{enumerate}
As can be seen from the protocol, the net key gain $\hat{G}$ per communication rounds of this protocol can be given by 
\begin{equation}
    \hat{G}=(\hat{N}_{\rm sift}-\hat{N}_{\rm KC} - \hat{N}_{\rm PA})/N. \label{eq:key_rate_finite}
\end{equation}
In Ref.~\cite{Boileau2005}, the protocol is shown to be secure in the finite-size regime by the so-called phase error correction approach \cite{Lo1999, Shor2000, Koashi2009}.  In fact, if we consider the purified version of the protocol defined above in which Alice instead prepares an entangled state $\ket{\Psi}_{AB}\coloneqq (\ket{0}_A\otimes R_y(2\pi\hat{r}_i/3)\ket{\psi_1}_B + \ket{1}_A\otimes R_y(2\pi(\hat{r}_{i} + 1)/3)\ket{\psi_1}_B)\sqrt{2}$, where $R_y(\theta)$ denotes the $\theta$ rotation around $y$ axis in the Bloch sphere, and sends the system $B$ to Bob, then the security of the key can be reduced to how much maximally entangled state $\ket{\Phi}_{AB}\coloneqq (\ket{0}_A\ket{0}_B + \ket{1}_A\ket{1}_B)/\sqrt{2}$ Alice and Bob can extract in this purified protocol.  Thus, the bit error operator $\Pi_{\rm bit}$ and phase error operator $\Pi_{\rm ph}$ can be defined, which correspond to the event that they obtain bit and phase flipped maximally entangled states, respectively, in the purified protocol.
Explicitly, $\Pi_{\rm bit}$ and $\Pi_{\rm ph}$ are given by \cite{Boileau2005}
\begin{align}
    \begin{split}
    &3\Pi_{\rm bit} \\
    &= \sum_{r=0}^{2}\Bigl[\ket{0}\!\bra{0}_A \otimes R_y\Bigl(\frac{2\pi r}{3}\Bigr)F\ket{1}\!\bra{1}_B F R_y\Bigl(-\frac{2\pi r}{3}\Bigr) \\
    &\quad  + \ket{1}\!\bra{1}_A \otimes R_y\Bigl(\frac{2\pi r}{3}\Bigr)F\ket{0}\!\bra{0}_B F R_y\Bigl(-\frac{2\pi r}{3}\Bigr)\Bigr],
    \end{split}\\
    \begin{split}
    & 3\Pi_{\rm ph} \\
    & = \sum_{r=0}^{2}\Bigl[\ket{+}\!\bra{+}_A \otimes R_y\Bigl(\frac{2\pi r}{3}\Bigr)F\ket{-}\!\bra{-}_B F R_y\Bigl(-\frac{2\pi r}{3}\Bigr) \\
    &\quad  + \ket{-}\!\bra{-}_A \otimes R_y\Bigl(\frac{2\pi r}{3}\Bigr)F\ket{+}\!\bra{+}_B F R_y\Bigl(-\frac{2\pi r}{3}\Bigr)\Bigr],
    \end{split}
\end{align}
where the filter operator $F\coloneqq \ket{0}\!\bra{0} + \frac{1}{\sqrt{3}}\ket{1}\!\bra{1}$.
From these, we can derive the following \cite{Boileau2005}:
\begin{equation}
    \Pi_{\rm ph} \leq \frac{5}{4}\Pi_{\rm bit}. \label{eq:ineq_bit_phase}
\end{equation}
Since the bit error rate $e_{\rm bit}$ can directly be observed in the ``test'' rounds, the asymptotic key rate $\hat{G}_{\rm aympt}$ of this protocol can be given by
\begin{equation}
    \hat{G}_{\rm asympt} = \frac{1}{4}\left(1 - h(e_{\rm bit}) - h\Bigl(\frac{5}{4}e_{\rm bit}\Bigr)\right),
\end{equation}
where $h(x)\coloneqq -x\log_2 x - (1 - x)\log_2(1 - x)$ denotes the binary-entropy function.   In the above, the prefactor $1/4$ comes from the fact that the measurement is conclusive with the probability $1/2$ (without Eve's attack) and that the ``sift'' is chosen with the probability $1/2$.  In the finite-size case, however, we need to take into account the statistical fluctuations.
Let us consider the purified protocol again and $\hat{x}_{\rm bit}^{(i)}$ be a random variable at $i$-th round that takes the value one when ``test'' is chosen and the bit error is observed while takes zero otherwise.  Then, $\hat{N}_{\rm bit}\coloneqq \sum_{i=1}^N \hat{x}_{\rm bit}^{(i)}$ corresponds to the number of bit error observed in the protocol. 
Let $\hat{x}_{\rm ph}^{(i)}$ be another random variable at $i$-th round that takes the value one when ``sift'' is chosen and the phase error occurs while takes zero otherwise.  Then, $\hat{N}_{\rm ph}\coloneqq \sum_{i=1}^N  \hat{x}_{\rm ph}^{(i)}$ corresponds to the number of phase errors in the sifted key.
It is known that if we can show
\begin{equation}
    \mathrm{Pr}[\hat{N}_{\rm ph} \geq U(\hat{N}_{\rm bit})] \leq \epsilon,  \label{eq:phase_error_bound}
\end{equation}
for a function $U(\hat{N}_{\rm bit})$ of $\hat{N}_{\rm bit}$, and by setting 
\begin{equation}
    \hat{N}_{\rm PA} = \hat{N}_{\rm sift} h(U(\hat{N}_{\rm bit})/ \hat{N}_{\rm sift}) + s, \label{eq:PA_shorten}
\end{equation}
then the protocol is $\sqrt{2(\epsilon + 2^{-s})}$-secret \cite{Koashi2009,Hayashi2012,matsuura2023digital}.  Furthermore, we assume that the secret-key consumption in the bit error correction at Step 4 of the protocol is given by 
\begin{equation}
    \hat{N}_{\rm KC} = \hat{N}_{\rm sift} h(\hat{N}_{\rm bit}/\hat{N}_{\rm sift}) + s',\label{eq:key_consumption_EC}
\end{equation}
for sending an $\hat{N}_{\rm sift} h(\hat{N}_{\rm bit}/\hat{N}_{\rm sift})$-bit syndrome string and $s'$-bit verification string to ensure $2^{-s'}$-correctness.  Thus, if we can obtain the function $U(\cdot)$ to satisfy Eq.~\eqref{eq:phase_error_bound}, we can obtain a key rate in the finite-size case as well satisfying $2^{-s'}$-correctness and $\sqrt{2(\epsilon + 2^{-s})}$-secrecy.

The derivation of Eq.~\eqref{eq:phase_error_bound} is very similar to the one in the previous section.
When Azuma's or Kato's is applied to the problem, the operator inequality~\eqref{eq:ineq_bit_phase} can be used as an inequality between the conditional expectations of $\hat{x}_{\rm bit}^{(i)}$ and $\hat{x}_{\rm ph}^{(i)}$.  By applying Azuma's inequality to $\hat{x}_{\rm ph}^{(i)} - \frac{5}{4}\hat{x}_{\rm bit}^{(i)}$, we have
\begin{equation}
    \mathrm{Pr}\left[\frac{\hat{N}_{\rm ph}}{N} - \frac{5}{4}\frac{\hat{N}_{\rm bit}}{N} \geq \left(1 + \frac{5}{4}\right)\sqrt{\frac{\ln(1 / \epsilon)}{2N}}\right] \leq \epsilon. \label{eq:phase_error_Azuma}
\end{equation}
In this case, the function $U_{\rm azuma}$ to satisfy Eq.~\eqref{eq:phase_error_bound} should be 
\begin{equation}
    U_{\rm azuma}(\hat{N}_{\rm bit}) = \frac{5}{4}\hat{N}_{\rm bit} + \frac{9}{4}\sqrt{\frac{N\ln(1 / \epsilon)}{2}}.\label{eq:upper_w_azuma}
\end{equation}
Similarly, by applying Kato's inequality to $\hat{x}_{\rm bit}^{(i)}$ while applying Azuma's inequality to $\hat{x}_{\rm ph}^{(i)}$, we have
\begin{align}
    \begin{split}
    &\mathrm{Pr}\Biggl[\frac{\hat{N}_{\rm ph}}{N} - \frac{5}{4}\frac{\hat{N}_{\rm bit}}{N} \\
    &\qquad \geq \sqrt{\frac{\ln(2 / \epsilon)}{2N}} + \frac{5}{4\sqrt{N}}\biggl[\beta - \alpha\Bigl(1 - \frac{2\hat{N}_{\rm bit}}{N}\Bigr)\biggr]\Biggr] \leq \epsilon,
    \end{split}\label{eq:phase_error_Kato} 
\end{align}
with
\begin{align}
    \alpha \coloneqq \alpha^{*}(N_{\rm bit}^{*},N,\epsilon/2),\\
    \beta \coloneqq \beta^{*}(N_{\rm bit}^{*},N,\epsilon/2),
\end{align}
where the functions $\alpha^*$ and $\beta^*$ are defined in Eqs.~\eqref{eq:alpha_star} and \eqref{eq:beta_star} with $N_{\rm bit}^{*}$ being an unconfirmed knowledge of $\hat{N}_{\rm bit}$ prior to the protocol.  The function $U_{\rm kato}$ to satisfy Eq.~\eqref{eq:phase_error_bound} should thus be
\begin{equation}
    \begin{split}
    U_{\rm kato}(\hat{N}_{\rm bit}) &= \frac{5}{4}\hat{N}_{\rm bit} + \sqrt{\frac{N \ln(2 / \epsilon)}{2}} \\
    &\qquad + \frac{5\sqrt{N}}{4}\left(\beta - \alpha\left(1-\frac{2\hat{N}_{\rm bit}}{N}\right)\right). 
    \end{split} \label{eq:upper_w_kato}
\end{equation}

Now we apply Theorem~\ref{theo:refined_de_finetti} to this protocol.  Let ${\cal X}$ be a sample space defined as 
\begin{equation}
    \begin{split}
    {\cal X} &\coloneqq \{\text{phase error in ``sift''}, \\
    & \qquad \text{bit error in ``test''},\, \text{others}\},
    \end{split}
\end{equation}
and $\bm{p}=(p_1, p_2, p_3)$ be a probability mass function on ${\cal X}$.  Then, from Eq.~\eqref{eq:ineq_bit_phase}, the set ${\cal Q}$ of allowed probability mass function for any quantum state between Alice and Bob can be given by
\begin{equation}
    {\cal Q} = \left\{\bm{p}=(p_1,p_2,p_3): p_1 \leq \frac{5}{4}p_2 \right\}, \label{eq:possible_prob_dist}
\end{equation}
where we used the fact that ``sift'' and ``test'' rounds are chosen with equal probability. 
Thus, for any i.i.d.~quantum state $\rho_{AB}^{\otimes N}$ between Alice and Bob, we have
\begin{equation}
    \begin{split}
    &\mathrm{Pr}\left[\left(\frac{\hat{N}_{\rm ph}}{N}, \frac{\hat{N}_{\rm bit}}{N}, 1-\frac{\hat{N}_{\rm ph}}{N}-\frac{\hat{N}_{\rm bit}}{N}\right)\in{\cal A}(\delta)\middle| \rho_{AB}^{\otimes N}\right]\\
    & \leq \max_{\bm{p}\in{\cal A}(\delta)}\max_{\bm{q}\in{\cal Q}}\exp[-N D(\bm{p}\|\bm{q})],
    \end{split}
\end{equation}
from Sanov's theorem \cite{Sanov1957,Csiszar1984}, where ${\cal A}(\delta)$ is defined as
\begin{equation}
    {\cal A}(\delta) = \left\{\bm{p}=(p_1,p_2,p_3): p_1 \geq \frac{5}{4}p_2 + \delta \right\}.
\end{equation}
Now, since the number of bit and phase errors is invariant under the permutation of rounds, we can virtually permute the four-dimensional quantum system between Alice and Bob in the purified protocol across the rounds.  Thus, we can apply Theorem~\ref{theo:refined_de_finetti} to the above and obtain
\begin{equation}
    \begin{split}
    &\mathrm{Pr}\left[\left(\frac{\hat{N}_{\rm ph}}{N}, \frac{\hat{N}_{\rm bit}}{N}, 1-\frac{\hat{N}_{\rm ph}}{N}-\frac{\hat{N}_{\rm bit}}{N}\right)\in{\cal A}(\delta)\right] \\
    &\leq \max_{\bm{p}\in{\cal A}(\delta)}\max_{\bm{q}\in{\cal Q}}f_q(N,4)\exp[-N D(\bm{p}\|\bm{q})].
    \end{split}
\end{equation}
We should optimize the value $\delta$ in the above so that the right-hand side is smaller than or equal to $\epsilon$.  If we write such an optimized $\delta$ in this case as $\delta_q$, the function $U_q$ to satisfy Eq.~\eqref{eq:phase_error_bound} is given by
\begin{equation}
    U_q(\hat{N}_{\rm bit}) = \frac{5}{4}\hat{N}_{\rm bit} + \delta_q. \label{eq:upper_w_qiid}
\end{equation}
Finally, we apply Theorem~\ref{theo:reduction_classical_iid} to this problem.  We can regard the measurement in the purified protocol as the three-outcome measurement channel as described above.  Since the set of probability mass functions after measurement is included in ${\cal Q}$ in Eq.~\eqref{eq:possible_prob_dist}, which is already linear and contains all the extremal points of the simplex of probability mass functions except $(1,0,0)$, we can choose ${\cal R}={\cal Q}$, where ${\cal R}$ is defined in Theorem~\ref{theo:reduction_classical_iid}.
The region ${\cal R}'$ in Theorem~\ref{theo:reduction_classical_iid} is thus given by
\begin{equation}
    \begin{split}
    {\cal R}' &= \Biggl\{\bm{p}=(p_1,p_2,p_3):\\
    &\hspace{1cm} p_1 - \frac{5}{4}p_2 \leq \frac{\sqrt{2}}{N}\sqrt{\left(\frac{13}{12}\right)^2 + \left(-\frac{14}{12}\right)^2 + \left(\frac{1}{12}\right)^2 }\Biggr\}.
    \end{split}
\end{equation}
Therefore, from Theorem~\ref{theo:reduction_classical_iid}, we have
\begin{equation}
    \begin{split}
    &\mathrm{Pr}\left[\left(\frac{\hat{N}_{\rm ph}}{N}, \frac{\hat{N}_{\rm bit}}{N}, 1-\frac{\hat{N}_{\rm ph}}{N}-\frac{\hat{N}_{\rm bit}}{N}\right)\in{\cal A}(\delta)\right]\\
    & \leq \max_{\bm{p}\in{\cal A}(\delta)}\max_{\bm{q}\in{\cal R}'}f_c(N,3)\exp[-N D(\bm{p}\|\bm{q})].
    \end{split}
\end{equation}
In the same way as the previous discussion, the value $\delta$ should be optimized so that the right-hand side above is smaller than or equal to $\epsilon$.  Let $\delta_c$ be such a value.  Then, the function $U_c$ to satisfy Eq.~\eqref{eq:phase_error_bound} in this case is given by
\begin{equation}
    U_c(\hat{N}_{\rm bit}) = \frac{5}{4}\hat{N}_{\rm bit} + \delta_c. \label{eq:upper_w_iim}
\end{equation}

\begin{figure}[t]
    \centering
    \includegraphics[width=0.98\linewidth]{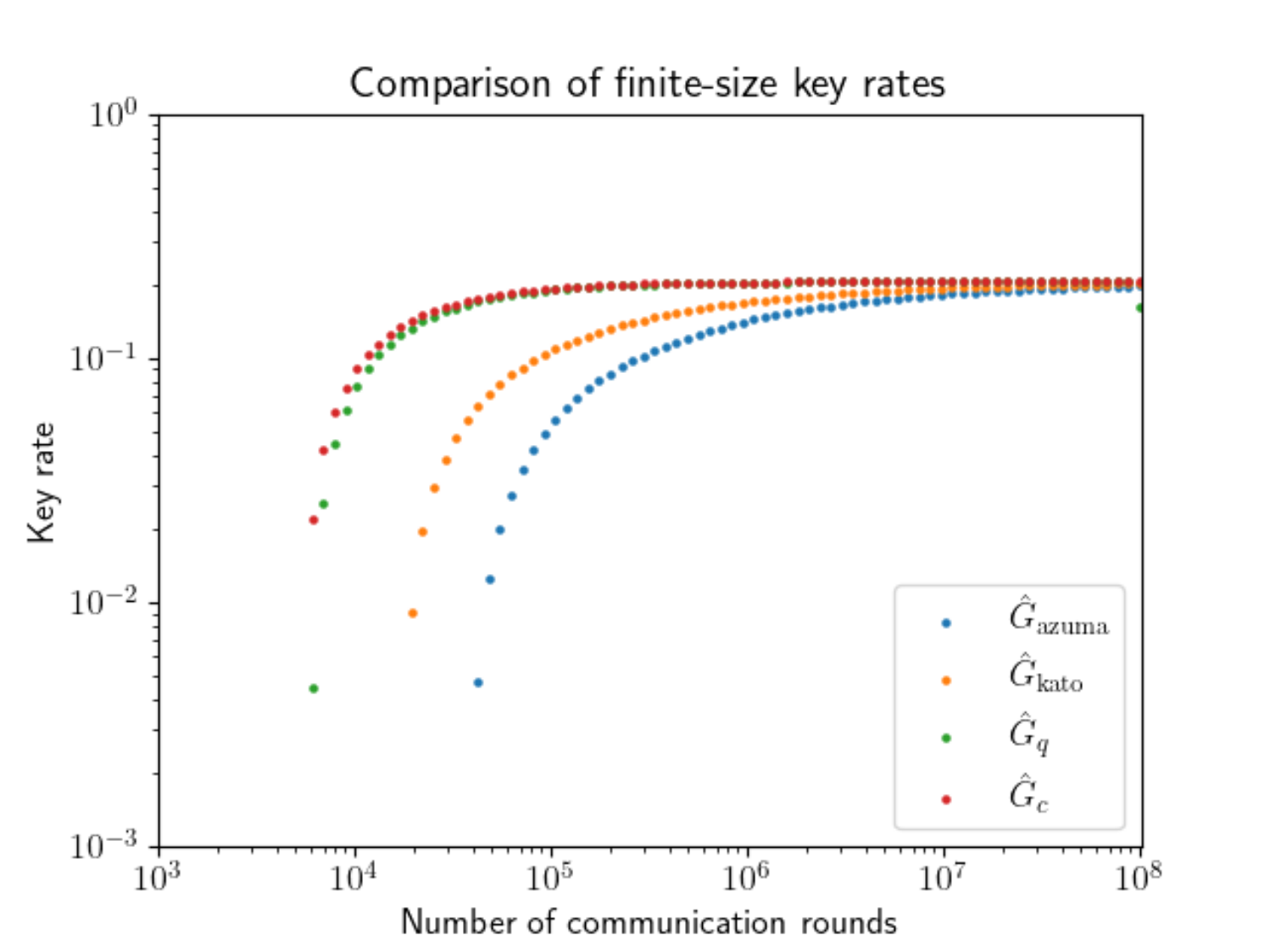}
    \caption{Comparison of key rates against the total number $N$ of rounds when the bit error rate is one percent.  The key rates are given by the formula~\eqref{eq:key_rate_formula}, where the function $U$ is either $U_{\rm azuma}$ in Eq.~\eqref{eq:upper_w_azuma}, $U_{\rm kato}$ in Eq.~\eqref{eq:upper_w_kato}, $U_q$ in Eq.~\eqref{eq:upper_w_qiid}, or $U_c$ in Eq.~\eqref{eq:upper_w_iim} with the parameter $\epsilon=2^{-102}$.  The parameters $s$ and $s'$ in Eq.~\eqref{eq:key_rate_formula} are chosen to be $s=102$ and $s'=50$ so that the correctness and secrecy parameter of the protocol are both $2^{-50}$.}
    \label{fig:key_rates}
\end{figure}

We compare the key rates obtained through the different concentration inequalities that give the functions Eqs.~\eqref{eq:upper_w_azuma}, \eqref{eq:upper_w_kato}, \eqref{eq:upper_w_qiid}, and \eqref{eq:upper_w_iim}, respectively.  We set $\epsilon=2^{-102}$, $s=102$, and $s'=50$ so that the secrecy and correctness parameters are $2^{-50}$, which are conventional values.
We also assume that the number $\hat{N}_{\rm con}$ of conclusive events is equal to its expectation value $N/2$ and that the length $\hat{N}_{\rm sift}$ of the sifted key is equal to its expectation value $N/4$.  The observed bit error rate is one percent, i.e., $\hat{N}_{\rm bit}=(N_{\rm con} - \hat{N}_{\rm sift})/100 = N/400$.  Under these assumptions, we computed the key rate $\hat{G}$ with respect to the total number $N$ of rounds, where $\hat{G}$ is given from Eqs.~\eqref{eq:key_rate_finite}, \eqref{eq:PA_shorten}, and \eqref{eq:key_consumption_EC} by:
\begin{equation}
    \hat{G} = \frac{\hat{N}_{\rm sift}[1 - h(\hat{N}_{\rm bit}/\hat{N}_{\rm sift}) - h(U(\hat{N}_{\rm bit})/\hat{N}_{\rm sift})] - s - s'}{N}. \label{eq:key_rate_formula}
\end{equation}
In the above, the function $U$ is either $U_{\rm azuma}$, $U_{\rm kato}$, $U_q$, or $U_c$.  
Figure~\ref{fig:key_rates} shows the comparison of the key rates against $N$ with the above four different concentration inequalities applied.  The figure shows that our concentration inequalities derived through Theorem~\ref{theo:refined_de_finetti} and \ref{theo:reduction_classical_iid} are better than those obtained through Azuma's and Kato's inequalities.  Therefore, also in this illustrated example, the Sanov-type quick convergence predominates over the polynomial factors in Eqs.~\eqref{eq:upper_w_qiid} and \eqref{eq:upper_w_iim}.

\section{Discussion}\label{sec:discussion}
In this paper, we developed concentration inequalities for a permutation-invariant quantum state possibly with an additional symmetry and a sequence of outcomes of independent and identical quantum measurements performed on an adversarial quantum state.  For the former case, we used the technique from the representation theory of the permutation group, which is well-known in the community of quantum information theory.  With this rather conventional technique, we can upper-bound the permutation-invariant state by i.i.d.~quantum states.  Compared to the previous de-Finetti-type statement \cite{Renner2008, Christandl2009, Fawzi2015}, our improvement for this case comes from the fact that one can adaptively choose these i.i.d.~states depending on the irreducible subspace of the permutation group.  

For the latter case, we found an upper bound on the probability of obtaining each sequence of quantum measurement outcomes with a specific type and then obtained a concentration inequality for the probability of obtaining sequences with the atypical types.
For this, we developed, to the best of the author's knowledge, a completely new technique that uses a geometric structure of the set of probability mass functions. 
Both bounds we obtained are Sanov-type and are thus expected to be tighter than the conventional bound obtained via Azuma's inequality.  In specific models we discussed in Sec.~\ref{sec:numerical_comparison}, our bounds even outperform the recently developed refinement of Azuma's inequality~\cite{Kato2020}.  

There remain a few questions on our results.  One of the questions is the range of applicability of our results compared to the conventional concentration inequalities such as Azuma's inequality.  As mentioned, our results require either permutation symmetry in quantum system or independent and identical measurements while Azuma's inequality does not.
There may be a few QKD protocols in which neither condition appears to hold such as the differential-phase-shift (DPS) protocol.  However, the state-of-the-art finite-size security proof of it decomposes the pulse train into blocks with three pulses to ease the analysis~\cite{Mizutani2023}, which may then make it possible to apply our results by actively imposing a permutation symmetry to the classical data.
Thus, detailed classification of the cases in which one can or cannot apply our results is left to a future work.

Another question is when our concentration inequalities outperform Kato's inequality~\cite{Kato2020}. From Fig.~\ref{fig:epsilon_vs_N}, it seems that our concentration inequalities can achieve smaller $\epsilon$ more quickly than Azuma's or Kato's inequality, and thus the use of our inequalities may be better in the case we need to ensure a small failure probability.  However, our concentration inequalities have a dependency either on the dimension of the quantum system (Theorem~\ref{theo:refined_de_finetti}) or the number of the measurement outcomes (Theorem~\ref{theo:reduction_classical_iid}) while Azuma's and Kato's inequalities do not.  Thus, which inequality performs better in a given setup is an open problem.

\begin{acknowledgments}
    Note that in the latest preprint, Nahar {\it et al.} considered another direction of refinement on the de Finetti theorem for a jointly permutation-symmetric extension of a fixed i.i.d.~reference state with and without local symmetry restrictions.  Their result reproduces the original post-selection technique~\cite{Christandl2009} as the special case when a reference system is trivial and there is no symmetry restriction.  As a result, a similar bound as Eqs.~\eqref{eq:concentration_with_symmetry} in this paper can be obtained through (independently studied) their result by setting the reference system trivial and taking a symmetry restriction into account. 
    This work was supported by the Ministry of Internal Affairs and Communications (MIC), R\&D of ICT Priority Technology Project (grant number JPMI00316); Council for Science, Technology and Innovation (CSTI), Cross-ministerial Strategic Innovation Promotion Program (SIP), ``Promoting the application of advanced quantum technology platforms to social issues'' (Funding agency:~QST); JSPS Overseas Research Fellowships; FoPM, WINGS Program, the University of Tokyo; JSPS Grant-in-Aid for Early-Career Scientists No.~JP22K13977.
\end{acknowledgments}

\bibliographystyle{quantum}
\bibliography{large_deviation.bib}
\end{document}